\documentclass[sigconf,natbib=true,screen=true]{acmart}

\acmSubmissionID{1227}

\usepackage{color}
\usepackage{bbm}
\usepackage{multirow}
\usepackage[inline]{enumitem}
\usepackage{graphicx}
\usepackage{subcaption}
\usepackage{sistyle}
\usepackage[ruled]{algorithm2e}
\usepackage{booktabs}
\usepackage{caption}
\SIthousandsep{,}
\usepackage{makecell}
\usepackage{amsmath}
\usepackage[english]{babel}
\usepackage{hyperref}
\usepackage{array} 
\usepackage{graphicx}
\usepackage{algorithmic}
\usepackage{tikz}
\usepackage{wrapfig}
\usepackage{acronym}
\newcommand{\heading}[1]{\vspace*{1mm}\noindent\textbf{#1.}}

\AtBeginDocument{%
  \providecommand\BibTeX{{%
    \normalfont B\kern-0.5em{\scshape i\kern-0.25em b}\kern-0.8em\TeX}}}

\makeatletter
\g@addto@macro\normalsize{%
  \abovedisplayskip 3pt plus1pt 
  \belowdisplayskip 3pt plus1pt
  \abovedisplayshortskip  0pt plus1pt%
  \belowdisplayshortskip  0pt plus1pt
}
\makeatother

\acrodef{CV}{computer vision}
\acrodef{IR}{information retrieval}
\acrodef{LLM}{large language model}
\acrodef{MDP}{Markov decision process}
\acrodef{NLP}{natural language processing}
\acrodef{NRM}{neural ranking model}
\acrodef{RL}{reinforcement learning}
\acrodef{RL-MARA}{Multi-grAnular Ranking Attack}
\acrodef{MoE}{mixture-of-experts}

\copyrightyear{2024}
\acmYear{2024}
\setcopyright{rightsretained}
\acmConference[SIGIR '24]{Proceedings of the 47th International ACM SIGIR Conference on Research and Development in Information Retrieval}{July 14--18, 2024}{Washington D.C., USA}
\acmBooktitle{Proceedings of the 47th International ACM SIGIR Conference on Research and Development in Information Retrieval (SIGIR '24), July 14--18, 2024, Washington D.C., USA}
\acmISBN{}
\acmDOI{}



\author{Yu-An Liu}
\orcid{0000-0002-9125-5097}
\affiliation{
	\institution{CAS Key Lab of Network Data Science and Technology, ICT, CAS}
	\institution{University of Chinese Academy of Sciences}
	\city{Beijing}
	\country{China}
}
\email{liuyuan21b@ict.ac.cn}

\author{Ruqing Zhang}
\orcid{0000-0003-4294-2541}
\affiliation{
	\institution{CAS Key Lab of Network Data Science and Technology, ICT, CAS}
	\institution{University of Chinese Academy of Sciences}
	\city{Beijing}
	\country{China}
}
\email{zhangruqing@ict.ac.cn}

\author{Jiafeng Guo}
\orcid{0000-0002-9509-8674}
\authornote{Jiafeng Guo is the corresponding author.}
\affiliation{
	\institution{CAS Key Lab of Network Data Science and Technology, ICT, CAS}
	\institution{University of Chinese Academy of Sciences}
	\city{Beijing}
	\country{China}
}
\email{guojiafeng@ict.ac.cn}

\author{Maarten de Rijke}
\orcid{0000-0002-1086-0202}
\affiliation{
 \institution{University of Amsterdam}
 \city{Amsterdam}
 \country{The Netherlands}
}
\email{m.derijke@uva.nl}

\author{Yixing Fan}
\orcid{0000-0003-4317-2702}
\affiliation{
	\institution{CAS Key Lab of Network Data Science and Technology, ICT, CAS}
 \institution{University of Chinese Academy of Sciences}
 \city{Beijing}
 \country{China}
}
\email{fanyixing@ict.ac.cn}

\author{Xueqi Cheng}
\orcid{0000-0002-5201-8195}
\affiliation{
	\institution{CAS Key Lab of Network Data Science and Technology, ICT, CAS}
	\institution{University of Chinese Academy of Sciences}
	\city{Beijing}
	\country{China}
}
\email{cxq@ict.ac.cn}

\renewcommand{\shortauthors}{Yu-An Liu et al.}

\begin{document}

\title[Multi-granular Adversarial Attacks against Black-box Neural Ranking Models]{Multi-granular Adversarial Attacks \\ against Black-box Neural Ranking Models}

\begin{abstract}
Adversarial ranking attacks have gained increasing attention due to their success in probing vulnerabilities, and, hence, enhancing the robustness, of neural ranking models. 
Conventional attack methods employ perturbations at a single granularity, e.g., word or sentence level, to target documents.  
However, limiting perturbations to a single level of granularity may reduce the flexibility of adversarial examples, thereby diminishing the potential threat of the attack. 
Therefore, we focus on generating high-quality adversarial examples by incorporating multi-granular perturbations. 
Achieving this objective involves tackling a combinatorial explosion problem, which requires identifying an optimal combination of perturbations across all possible levels of granularity, positions, and textual pieces. 
To address this challenge, we transform the multi-granular adversarial attack into a sequential decision-making process, where perturbations in the next attack step build on the perturbed document in the current attack step. 
Since the attack process can only access the final state without direct intermediate signals, we use reinforcement learning to perform multi-granular attacks. 
During the reinforcement learning process, two agents work cooperatively to identify multi-granular vulnerabilities as attack targets and organize perturbation candidates into a final perturbation sequence. 
Experimental results show that our attack method surpasses prevailing baselines in both attack effectiveness and imperceptibility. 
\end{abstract}

\maketitle

\section{Introduction}
With the advance of deep neural networks \cite{lecun2015deep}, \acp{NRM} \citep{ZhuyunDai2019DeeperTU,guo2016deep,mitra2017learning,KezbanDilekOnal2018NeuralIR} have achieved promising ranking effectiveness in \ac{IR}. 
Besides their proven effectiveness, considerable attention has been directed toward assessing the adversarial robustness of \acp{NRM}.

\heading{Adversarial ranking attacks}
In IR, \acp{NRM} are prone to inheriting vulnerabilities to adversarial examples from general neural networks~\cite{raval2020one,wu2022prada,liu2022order}.
Such adversarial examples are crafted by introducing human-imperceptible perturbations to the input, capable of inducing model misbehavior. 
This discovery has sparked legitimate concerns about potential exploitation by black-hat SEO practitioners aiming to defeat meticulously designed search engines \cite{gyongyi2005web}. 
Consequently, there is a need to develop robust and reliable neural \ac{IR} systems. 
A crucial step in this direction involves introducing \textit{adversarial ranking attacks} to benchmark the vulnerability of black-box \acp{NRM} 
\cite{wu2022prada,liu2022order,chen2023towards}. 
Here, the aim of an adversary is to find human-imperceptible perturbations injected into the document's text, to promote a low-ranked document to a higher position in the ranked list produced for a given query \citep{wu2022prada,chen2023towards}. 
This attack approach allows us to identify vulnerabilities in \acp{NRM} before deploying them in real-world settings and devise effective countermeasures. 

\heading{Single-granular ranking attacks}
Existing studies on adversarial attacks against \acp{NRM} are typically restricted to document perturbation strategies that operate at a single level of granularity, such as the word-level~\citep{raval2020one,wu2022prada} or sentence-level \cite{liu2022order,chen2023towards}. 
These methods face two main limitations: 
\begin{enumerate*}[label=(\roman*)]
\item \textit{Different perturbation granularities for different query-document pairs}: 
When dealing with different query-document pairs, \textit{a priori} limiting perturbations to a single granularity could considerably restrict the choice of attack targets, thereby impeding the overall effectiveness of the attack.
\item \textit{Multiple perturbation granularities for a query-document pair}: 
Despite employing a meticulously chosen attack granularity tailored to specific query-document pairs, a fixed granularity falls short of fully capturing the diverse relevance patterns inherent in the matching between a query and a document \cite{fan2018modeling}.
An effective ranking attack should possess the flexibility to simultaneously consider various granularity perturbations in an adversarial example. 
\end{enumerate*}
In this sense, we argue that the full potential of adversarial attacks has yet to be harnessed for uncovering the vulnerabilities of \acp{NRM}.

\begin{figure}[t]
    \centering
    \includegraphics[width=\linewidth]{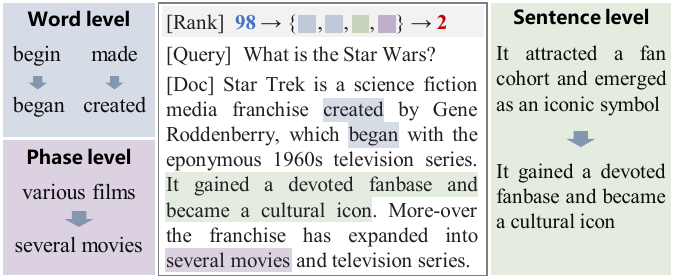}
    \caption{To prompt a target document in the rankings to a query, we identify multi-granular texts within the document as attack targets to generate effective adversarial examples.}
    \label{fig:pertubation type}
\end{figure}

\heading{Multi-granular adversarial ranking attacks}
In this paper, we develop multi-granular adversarial ranking attacks against \acp{NRM}; see Figure~\ref{fig:pertubation type} for an illustrative example. 
We incorporate word-level, phrase-level, and sentence-level perturbations to generate fluent and imperceptible adversarial examples. 
The generated examples may not cover all three levels of granularity but allow for flexible selection based on an optimization strategy. 
Compared to existing single-granular attacks, this multi-granular approach broadens the selection of attack targets and explores the vulnerability distribution of NRMs at various granularities within a specific query-document pair. 
Consequently, it can yield richer and more diverse forms of adversarial examples, thereby enhancing campaign performance.

\heading{Learning multi-granular attack sequences}
Achieving a multi-granular attack is non-trivial due to the combinatorial explosion arising from the numerous possible actions, e.g., perturbation granularities, target positions in the document, and replacement content, posing a significant computational challenge. 
To address this challenge, we formulate the multi-granular ranking attack problem as a sequential decision-making process \cite{mazyavkina2021reinforcement, nguyen2017collective}. 
In this process, the attacker sequentially introduces a perturbation at a specific level of granularity, i.e., word-level, phrase-level, or sentence-level, guided by the perturbations in the preceding steps.

Within the sequential decision-making process, the discrete and non-differentiable nature of the text space presents a challenge in finding a direct supervisory signal to facilitate the incorporation of multi-granular perturbations. 
Therefore, we propose \acs{RL-MARA}, a novel \underline{r}einforcement \underline{l}earning (\acs{RL}\acused{RL}) framework \cite{sutton2018reinforcement} to navigate an appropriate sequential \underline{m}ulti-gr\underline{a}nular \underline{r}anking \underline{a}ttack\acused{RL-MARA} path. 

Following~\citep{wu2022prada, liu2022order, chen2023towards}, our focus is on a practical and challenging decision-based black-box setting \cite{brendel2021decision}, where the adversary lacks direct access to model information and can only query the target \ac{NRM}. 
We train a surrogate ranking model to substitute and achieve comparable performance to the black-box target \ac{NRM}. 
We combine the surrogate ranking model with a \ac{LLM} to form the complete environment, to provide rewards for assessing the effectiveness of the multi-granular attack and the naturalness of the perturbed document, respectively. 
We set up a multi-granular attacker by building upon existing single-granular attack methods agents. 
A sub-agent and a meta-agent are designed in a cooperative manner: the sub-agent is tasked to identify multi-granular vulnerabilities in the document as attack targets, while the meta-agent is tasked to generate and organize perturbations into a final perturbation sequence, respectively. 
During the \ac{RL} process, the attacker sequentially incorporates perturbations until the cumulative perturbation exceeds a predefined budget. 

\heading{Main findings}
We conduct experiments on two web search benchmark datasets, MS MARCO Document Ranking~\cite{nguyen2016ms} and ClueWeb09-B~\cite{clarke2009overview}. 
Experimental results demonstrate that \ac{RL-MARA} significantly improves the document ranking of target documents and thus achieves a higher attack success rate than existing single-granular ranking attack baselines.
According to automatic and human naturalness evaluations, \ac{RL-MARA} could maintain the semantic consistency and fluency of adversarial examples.

\vspace{-2mm}
\section{Problem Statement}
In ad-hoc retrieval, given a query $q$ and a set of $N$ document candidates $\mathcal{D} = \{d_1,d_2,\ldots,d_N\}$ from a corpus $\mathcal{C}$, the objective of a ranking model $f$ is to assign a relevance score $f(q,d_n)$ to each pair of $q$ and $d_n \in \mathcal{D}$, to obtain the ranked list $L$. 


\heading{Adversarial ranking attack} Many studies have examined adversarial ranking attacks against NRMs \cite{wu2022prada,liu2022order,liu2023topic,chen2023towards}. 
Given a target document $d$ and a query $q$, the primary goal is to construct a valid adversarial example $d^\mathrm{adv}$ capable of being ranked higher than the original $d$  in response to $q$ by NRMs, all while closely resembling $d$. 
The adversarial example $d^\mathrm{adv}$ can be regarded as $d \oplus \mathcal{P}$, where $\mathcal{P}$ denotes the perturbation applied to $d$. 
The perturbations $\mathcal{P}$ are crafted to conform to the following properties~\citep{wu2022prada,chen2023towards,liu2022order}:
\begin{equation}
\label{eq:form}
\begin{split}
 &\operatorname{Rank}(q,d \oplus \mathcal{P}) < \operatorname{Rank}(q,d)  \\
 &\quad \text{ such that } \| \mathcal{P} \| \leq \epsilon, \ \operatorname{Sim}(d^\mathrm{adv}, d) \geq \lambda,
\end{split} 
\end{equation}
where $\operatorname{Rank}(q,d \oplus \mathcal{P})$ and $\operatorname{Rank}(q,d)$ denote the position of $d^\mathrm{adv}$ and $d$ in the ranked list with respect to $q$, respectively;
a smaller value of ranking position denotes a higher ranking;
$\epsilon$ represents the budget for the number of manipulated terms $\| \mathcal{P} \|$;
$\lambda$ is the coefficient; and the function $\operatorname{Sim}(d^\mathrm{adv}, d)$ assesses the semantic or syntactic similarity \cite{wu2022prada,fang-etal-2023-modeling} between $d$ and its corresponding $d^\mathrm{adv}$.
Ideally, $d^\mathrm{adv}$ should preserve the original semantics of $d$,  and be imperceptible to human judges yet misleading to NRMs. 

\heading{Decision-based black-box attacks} 
Following~\cite{wu2022prada,liu2022order}, we focus on decision-based black-box attacks against NRMs for the adversarial ranking attack task. 
This choice is motivated by the fact that the majority of real-world search engines operate as black boxes, granting adversaries access only to the final decision, i.e., the rank positions within the partially retrieved list.

\heading{Perturbations at multiple levels of granularity} 
Existing work mainly focuses on a single granularity of $\mathcal{P}$ to manipulate the target document, esp.\ word-level word substitution \cite{wu2022prada} and sentence-level trigger generation  \cite{liu2022order}. 
However, restricting perturbations to a single granularity may fail to adequately capture the nuanced and diverse vulnerability features, a limitation confirmed by our experimental results (see Section \ref{Sec: Attack evaluation}). 
Thus, we propose to find perturbations $\mathcal{P}$ at multiple levels of granularity. 

Effective character-level \citep{zhang2020adversarial} and phrase-level \citep{lei2022phrase} modifications have proven successful in textual attacks \cite{zhang2020adversarial} within NLP, but are underutilized in IR. 
However, character-level attacks tend to create ungrammatical adversarial examples and are easily defended against \cite{pruthi2019combating}. 
And considering the naturalness requirements of the adversarial examples, introducing perturbations at higher levels of granularity, e.g., paragraph level, may pose challenges in avoiding suspicion. 
Therefore, we propose to launch an adversarial ranking attack at three levels of perturbation granularity, i.e., word, phrase and sentence levels. 
\vspace*{-4.5mm}
\section{Preliminaries} \label{Sec:Pre}
Our approach relies on three single-granular adversarial ranking attack methods, i.e., word-level, phrase-level and sentence-level. 
Typically, single-granular attacks begin by training a surrogate ranking model that imitates the target \ac{NRM}. Subsequently, they execute attacks guided by the surrogate model, which involve two primary steps:
\begin{enumerate*}[label=(\roman*)]
    \item Identifying vulnerable positions in the target document; and
    \item Perturbing the text at these identified positions.
\end{enumerate*}

\heading{Word-level attack} 
For word-level attacks against \acp{NRM}, the main approaches include word substitution \cite{wu2022prada}, word insertion \cite{wang2022bert}, and word removal \cite{samanta2017towards}.
In this study, we employ the word substitution ranking attack exemplified by PRADA \cite{wu2022prada}, which 
has shown promising results in terms of the attack success rate. 
Specifically, PRADA first identifies vulnerable words in a document that significantly influence the final ranking result through the surrogate model; and then replaces these vulnerable words with synonyms, selecting the one that provides the most substantial boost in rankings from a pool of candidate synonyms. 

\heading{Phrase-level attack} In textual attacks within \ac{NLP}, the predominant method for phrase-level attacks \cite{zheng2020evaluating,lei2022phrase} is phrase substitution \cite{lei2022phrase}. 
To the best of our knowledge, there has been an absence of phrase-level attacks specifically targeting \acp{NRM} in \ac{IR}. 
PLAT \cite{lei2022phrase} stands out in phrase-level textual attacks, aiming to induce text misclassification. 
PLAT first identifies the positions of vulnerable phrases that significantly influence the classification scores predicted by the surrogate model.
Then, it utilizes BART \cite{lewis2020bart} to generate multiple variations for each selected vulnerable phrase.
PLAT selects the variant that introduces the most substantial interference in classification scores predicted by the surrogate model, deviating from the original phrase.
To adapt PLAT from classification to ranking, we use a sub-agent (see Section~\ref{Sec: Sub-agent}) to find positions of important phrases and replace classification scores with relevance scores.

\heading{Sentence-level attack} 
For sentence-level attacks against \acp{NRM}, key strategies encompass sentence substitution \cite{liu2022order}, sentence insertion \cite{chen2023towards}, and sentence rewriting \cite{song2022trattack}.
Here, we employ the sentence substitution ranking attack exemplified by PAT \cite{liu2022order}, which replaces a sentence at a specific position in a document with a trigger. 
Specifically, PAT first designates the beginning of the document as the vulnerable sentence; and then optimizes the gradients of the ranking loss to derive a continuous trigger representation, which is mapped to the word space. 
In this work, we take a more flexible way, employing the sub-agent to identify important sentences, potentially located anywhere within the document.

\heading{Combining single-granular attacks}
In our multi-granular attack:  
\begin{enumerate*}[label=(\roman*)]
\item Initially, a sub-agent, serving as a vulnerability indicator (see Section~\ref{Sec: Sub-agent}), is employed to identify important positions at each level of granularity.  
\item Then, the three aforementioned single-granular attack methods are applied to generate a perturbation for each identified important position (see Section~\ref{Sec: meta-agent}). 
\item Finally, an organization of all settled perturbations is executed to refine and select the most effective perturbation sequence. 
\end{enumerate*}

\heading{Discussions} 
Our framework can seamlessly integrate off-the-shelf ranking attack methods.
In this work, for each granularity level, we choose a representative attack method, rather than involve several attack methods. 
Involving more attack methods at the same granularity may introduce additional variables and complexities, potentially confounding our results and making it more challenging to draw clear conclusions. 
In the future, we plan to consider more granularity and add more attack methods at the same granularity to make the perturbation even more diverse.
Besides, the above three single-granular attacks remain constant in our current multi-granular attack method. 
In the future, we aim to make these attack methods learnable and dynamically update them within the entire framework to achieve enhanced interoperability. 

\vspace{-2mm}
\section{Method}
In this section, we introduce the \ac{RL-MARA} framework, specifically crafted for achieving multi-granular attacks against \acp{NRM}. 

\begin{figure}[t]
    \centering
    \includegraphics[width=\linewidth]{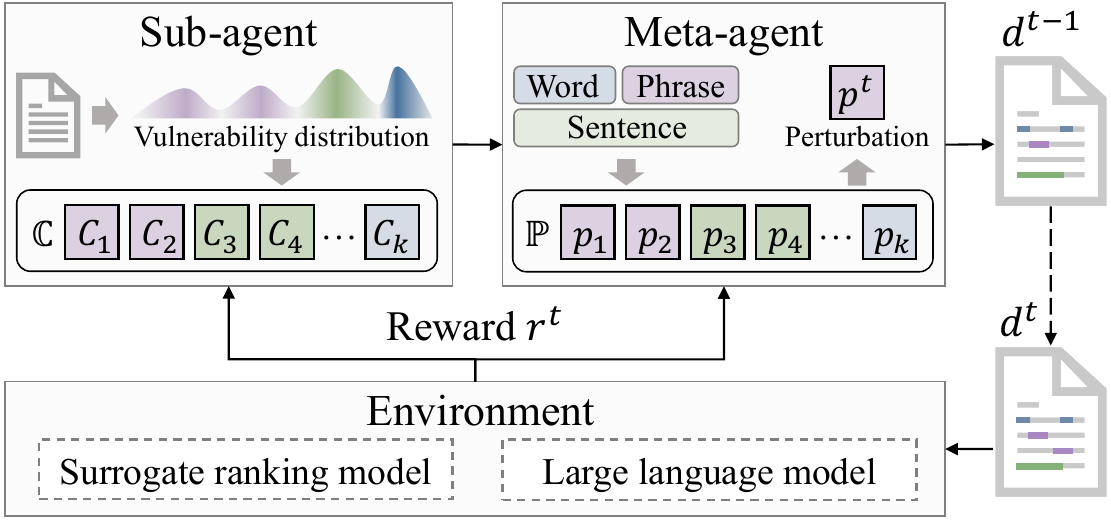}
    \caption{The \ac{RL-MARA} framework.}
    \label{fig:framework}
    \vspace{-2mm}
\end{figure}

\vspace{-2mm}
\subsection{Overview}
To generate multi-granular perturbations for a target document, we need to decide the granularity, position, and modified content of each single perturbation.
We formulate multi-granular attacks against target \acp{NRM} as a sequential decision-making process: 
\begin{enumerate*}[label=(\roman*)]
\item The attackers manipulate the target document by introducing a perturbation that could be at any level of granularity; a surrogate model of the target \ac{NRM} assesses the current ranking position, while a \acf{LLM} evaluates its naturalness; 
\item The attacker observes changes in ranking and naturalness and further optimizes its attack strategy to generate the next perturbation. 
\end{enumerate*}  
The global objective is to optimize the final ranking improvement of the target document with indiscernible perturbations.

During the sequential decision-making process, the discrete perturbed document leads to a lack of direct signals at each step.
Therefore, we employ \acl{RL} (\ac{RL}) to identify an appropriate sequential attack path for generating adversaries. 
Specifically, we introduce the RL-based framework \ac{RL-MARA} to learn an optimal multi-granular ranking attack strategy. 
As shown in  Figure~\ref{fig:framework}, the \ac{RL-MARA} framework comprises two major components: 
\begin{enumerate*}[label=(\roman*)]
\item A surrogate model simulating the behavior of the target \ac{NRM}, and an advanced \ac{LLM} evaluating the naturality of adversarial samples, collectively serve as the whole environment to provide rewards; and 
\item A multi-granular attacker, consisting of a sub-agent and a meta-agent, receives rewards from the environment and collaborates to generate perturbations at multiple levels of granularity. 
\end{enumerate*}

\vspace{-4mm}
\subsection{Environment and reward}
The multi-granular ranking attack problem is formally modeled as a Markov decision process (\ac{MDP}) \cite{bellman1957markovian}, wherein its components are defined as follows:  
\begin{enumerate*}[label=(\roman*)]
\item \textbf{State} $d$ is the document, with the initial state $d^0$ as a target document, and the terminal state signifying a successful adversarial example;
\item \textbf{Action} $p$ denotes a multi-granular perturbation selected by the agent for injection into the document;
\item \textbf{Transition} $\mathcal{T}$ alters the document state $d$ by applying a perturbation at each step; and
\item \textbf{Reward} $r$ is the attack reward provided by the environment, guiding the agent with supervisory signals.
\end{enumerate*}

\heading{Environment} 
In the decision-based black-box scenario, obtaining only hard-label predictions and lacking the relevance score for each candidate document predicted by the target \ac{NRM}, poses a challenge. 
Besides, frequent queries to the target \ac{NRM} might arouse suspicion.
Consequently, we employ a surrogate ranking model to function as an environment and offer the attack reward as a proxy for the target \ac{NRM}.  
Following \cite{wu2022prada,liu2022order}, we train a surrogate ranking model based on the Pseudo Relevance Feedback idea \cite{dehghani2017neural} and achieve comparable performance to the target \ac{NRM}.
Specific training details can be found in Section \ref{sec: details}. 
Simultaneously, we introduce an advanced \ac{LLM} as part of the environment to assess the naturalness of the current perturbed document as a reward. 
To sum up, the virtual environment for the RL Attacker comprises a surrogate ranking model and an \ac{LLM}. 

\heading{Multi-granular reward design} 
An effective reward function for multi-granular attacks should consider both attack effectiveness and the naturalness of the perturbed document. 
At each step of the sequential interactions, the attacker introduces a perturbation at a specific level of granularity. 
The reward furnishes appropriate feedback based on the granularity of the current perturbation, guiding the behavior of the attacker. 

Specifically, the reward for each step is defined as follows: 
\begin{enumerate*}[label=(\roman*)] 
\item If the relevance score of the current perturbed document is higher than before, the current attack \textit{succeeds}. 
The reward not only evaluates the attack effectiveness and the naturalness of the perturbed document, but also considers the number of manipulated terms introduced by different perturbation granularities; and  
\item Conversely, if the attack \textit{fails}, we directly apply a fixed penalty factor $\xi$ as the reward. 
\end{enumerate*}
These assumptions lead us to define the attack reward function $r^t$ at the step $t$ as follows: 
\begin{equation}
r^t =\left\{
\begin{array}{ll}
-\xi, & \text{ if } \tilde{f}(q,d^{t}) < \tilde{f}(q,d^{t-1}) \\
r^t_\mathrm{att} / | p^t | + \beta r^t_\mathrm{nat} , &  \text{else,}  \\ 
\end{array}
\right.
\end{equation}
where $d^{t}$ and $d^{t-1}$ are the perturbed document at step $t$ and $t-1$, respectively; 
$r^t_\mathrm{att}$ and $r^t_\mathrm{nat}$ are the rewards with respect to attack effectiveness and document naturalness, respectively. 
The penalty factor $\xi$ is set to 1.
The function $\tilde{f}\left( \cdot \right)$ outputs the relevance score judged by the surrogate ranking model $\tilde{f}$, and the hyper-parameter $\beta$ balances attack effectiveness with document naturalness. 

To emphasize the impact of different levels of granularity, we introduce $| p^t |$ as a reward discount of attack effectiveness, representing the number of manipulated terms of $p^t$. 
The value of $| p^t |$ varies significantly across perturbations at different levels of granularity. 
For instance, when an attacker introduces a sentence-level perturbation, it consumes a larger portion of the perturbation budget. To normalize its effect, given its expected stronger attack effects, we incorporate a corresponding discount factor.

Next, we detail $r^t_\mathrm{att}$ and $r^t_\mathrm{nat}$:  

 \begin{figure}[t]
    \centering
    \includegraphics[width=\linewidth]{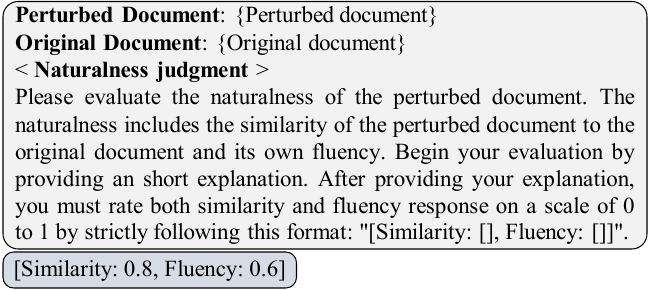}
    \caption{Instruction for naturalness evaluation with chatGPT. The gray and dark blue blocks indicate the inputs and outputs of the model, respectively.}
    \label{fig: prompt}
    \vspace{-1mm}
\end{figure}

\begin{itemize}[leftmargin=*]
\item \textbf{Attack effectiveness reward.} 
$r^t_\mathrm{att}$ incentivizes ranking improvements of the perturbed document $d^t$; a perturbed document should receive more rewards if it is ranked higher than before. 
However, directly using ranking as a reward is sparse.   
We shape the reward using the surrogate model's relevance scores, i.e.,
\begin{equation}
r^t_\mathrm{att} = \tilde{f}(q,d^{t})-\tilde{f}(q,d^{t-1}),
\end{equation}
where $\tilde{f}\left( \cdot \right)$ outputs the relevance score judged by $\tilde{f}$.
If the step $t$ attack is successful, $r^t_\mathrm{att}$ is positive.

\item \textbf{Document naturalness reward.} $r^t_\mathrm{nat}$
guarantees that the perturbed document $d^t$ satisfies semantic and syntactic constraints by an \ac{LLM} \cite{ni2024llms,li2023generative,he2023merging}.
In this work, we employ ChatGPT \cite{chatgpt} as the \ac{LLM}.
Specifically, we use the Prompts, shown in Figure~\ref{fig: prompt}, to evaluate both the similarity and fluency of documents in each state, defined as  $r^t_\mathrm{sim}$ and  $r^t_\mathrm{flu}$, respectively. 

\begin{itemize}[label={\scriptsize$\blacksquare$}]
    \item $r^t_\mathrm{sim}$ measures how semantically similar the perturbed document $d^t$ is to the original document:
    \begin{equation}
        r^t_\mathrm{sim} = \operatorname{LLM_\mathrm{sim}}\left(d^{t},d^{0}\right),
    \end{equation}
    where the function $\operatorname{LLM_\mathrm{sim}}\left( \cdot \right)$ outputs the similarity score judged by the \ac{LLM}.
    \item $r^t_\mathrm{flu}$ measures how fluent the perturbed document $d^t$ is:
    \begin{equation}
        r^t_\mathrm{flu} = \operatorname{LLM_\mathrm{flu}}\left(d^{t}\right),
    \end{equation}
    where the function $\operatorname{LLM_\mathrm{flu}}\left( \cdot \right)$ outputs the fluency score judged by the \ac{LLM}.
\end{itemize}
Finally, the overall reward with respect to document naturalness $r^t_\mathrm{nat}$ is defined as $r^t_\mathrm{nat} = r^t_\mathrm{sim} + r^t_\mathrm{flu}$. 
\end{itemize}

\vspace{-2mm}
\subsection{Multi-granular attacker}
The objective of the multi-granular attacker is to identify possible attack positions at all granularities and organize them into a final perturbation sequence. 
Developing such a composite strategy is challenging for a single agent.
Consequently, we adopt two agents in a cooperative manner to accomplish this goal. 
\begin{enumerate*}[label=(\roman*)]
\item The \textbf{vulnerability indicator} serves as a sub-agent, aiming to identify all vulnerable positions in the document at each level of perturbation granularity. 
\item The \textbf{perturbation aggregator} acts as a meta-agent, aiming to generate specific perturbations for each selected vulnerable position and organize them to filter out the final perturbation sequence.
\end{enumerate*}

\subsubsection{Sub-agent: vulnerability indicator}
\label{Sec: Sub-agent}
The aim of the vulnerability indicator is to identify all vulnerable positions within the target document and the corresponding level of perturbation granularity of each position.

\heading{Policy network}
We employ BERT \cite{devlin2018bert} as the backbone of the sub-agent policy network $I_{\phi}$. 
Given a target document $d$ and a query $q$, the process proceeds as follows: 
\begin{itemize}[leftmargin=*]
\item We employ the surrogate model $\tilde{f}$ to compute the pairwise loss 
$ \mathcal{L}_\mathrm{pair} = \sum_{d^{\prime} \in L_{\backslash d}} \mathcal{L}_{\tilde{f}}\left(q, d, d^{\prime}\right)$, 
where $d^{\prime}$ is the remaining documents in the ranked list $L$ excluding the target document $d$.

\item We compute the average gradient $\boldsymbol{g}_{d} \in \mathbb{R}^{m * l} $ of $\mathcal{L}_\mathrm{pair}$ concerning each position (i.e., each word) in the target document $d$.
Here, $m$ represents the dimensions of the hidden state, and $l$ denotes the length of the target document $d$. 

\item We feed $\boldsymbol{g}_{d}$ into the vulnerability indicator $I_{\phi}$, to derive the vulnerability distribution $\boldsymbol{u}$.  
The vulnerability distribution $\boldsymbol{u}$ represents a list of confidence scores indicating the confidence level for each position in the target document across various levels of perturbation granularity, which is calculated by:
\begin{equation}
\boldsymbol{u} = I_{\phi} \left(\boldsymbol{g}_{d}\right),
\end{equation}
where $\boldsymbol{u} = \left\{\boldsymbol{u}_1,\boldsymbol{u}_2,\ldots,\boldsymbol{u}_l\right\} \subseteq \mathbb{R}^{4*l}$. Each $\boldsymbol{u}_i \in \mathbb{R}^4, i \in \left[1,l\right]$, represents the confidence scores at each level of perturbation granularity at position $i$. The four dimensions correspond to perturbation at word-level (W), phrase-level (P), sentence-level (S), and no perturbation (N), respectively.

\item To condense the vulnerability distribution into a specific perturbation type, we apply the softmax function to $\boldsymbol{u}$, yielding the vulnerable word positions $\boldsymbol{c}$ for the target document $d$, i.e., 
\begin{equation}
\boldsymbol{c} = \operatorname{softmax}\left(\boldsymbol{u}\right),
\end{equation}
where $\boldsymbol{c} = \left\{c_1,c_2,\ldots,c_l\right\} \subseteq \mathbb{R}^{1 * l}$, and each $c_i \in \{\operatorname{W}, \operatorname{P}, \operatorname{S},\operatorname{N}\} , i \in \left[1,l\right]$, is the granularity of perturbation at each word position $i$. 
\end{itemize}

\noindent%
During the above process, to ensure the continuity of vulnerable positions at the phrase and sentence granularity, we constrain the output of the vulnerability indicator as part of a sequence labeling process, as detailed in Section~\ref{sec: details}.
Based on this, we can map the vulnerable word positions to corresponding \textit{span positions} at the word, phrase, and sentence levels, denoted as $\mathbb{C} =  \left\{C_1,C_2,\ldots,C_k\right\} \in \mathbb{R}^{1 * k}, k < l$.
Each $C_j \in \{\operatorname{W}, \operatorname{P}, \operatorname{S}\}, j \in \left[1,k\right]$, represents the level of perturbation granularity at each span $j$.

\vspace{-2mm}
\subsubsection{Meta-agent: perturbation aggregator} 
\label{Sec: meta-agent}
The target of the perturbation aggregator is to generate specific perturbations for each selected vulnerable span position and organize them into a final perturbation sequence.

\heading{Generating specific perturbations for each selected span position via static single-granular attack methods}
Once we have identified the vulnerable span positions with the corresponding perturbation granularity $\mathbb{C}$ with the vulnerability indicator, we employ the respective attack method (outlined in Section~\ref{Sec:Pre}) for each of these span positions to generate the specific perturbations $\mathbb{P} \in \mathbb{R}^k$. 
Based on $\mathbb{P}$, we design the policy network, which sequentially selects a perturbation $p^t$ from  $\mathbb{P}$, adding to the target document until the budget of manipulated words number $\epsilon$ is reached.

\heading{Policy network} 
We employ a multi-layer perception (MLP) \cite{rumelhart1985learning} as the backbone of the meta-agent policy network $G_{\varphi}$.
For each step $t$, $G_{\varphi}$ takes the perturbed document $d^{t-1}$ and vulnerability distribution $\boldsymbol{u}$ as inputs.
The action $p^t$ is to select the $t$-th specific perturbation added to the document.
The process is as follows: 
\begin{itemize}[leftmargin=*]
\item We use the surrogate model $\tilde{f}$ to obtain the hidden states of the perturbed document $d^{t-1}$ as $\boldsymbol{h}^{t-1} = [\boldsymbol{h}^{t-1}_1, \boldsymbol{h}^{t-1}_2, \ldots, \boldsymbol{h}^{t-1}_l]$, where $l$ is the length of $d^{t-1}$. 
$\boldsymbol{h}^{t-1}_i \in \mathbb{R}^m $ is the hidden state of the $i$-th word in $d^{t-1}$, where $m$ is the dimension of hidden states.

\item For each span (e.g., word, phrase, sentence), let's consider the $j$-th span. From its starting position $j\text{-start}$ to its ending position $j\text{-end}$, we concatenate the corresponding hidden state with the confidence score. 
Then, we sum them up to create the concatenated representation, denoted as, 
$\sum_{o = j,start}^{j,end}\left[\boldsymbol{h}^{t-1}_o ; \boldsymbol{u}_o\right]$, where $\left[; \right]$ is the concatenation operation.

\item We divide the concatenated representation by the length of the span to derive the final representation $\boldsymbol{e}_j^{t-1}$ for the current state:  
\begin{equation}
\boldsymbol{e}_j^{t-1}  = \left(\sum\nolimits_{o = j,start}^{j,end}\left[\boldsymbol{h}^{t-1}_o ; \boldsymbol{u}_o\right]\right) / \left|p_j\right|, j \in \left[1,k\right],
\end{equation}
where $\left|p_j\right|$ is length of the specific perturbation of $j$-th span in $\mathbb{P}$.

\item The probability distribution of each specific perturbation $p(p_j\mid d^{t-1})$ at $t$-step can be calculated by perturbation aggregator $G_{\varphi}$:
\begin{equation}
P(p_j\mid d^{t-1})  = G_{\varphi}(\boldsymbol{e}_j^{t-1}), \ p_j \in \mathbb{P}. 
\end{equation}
The aggregator samples the perturbation $p^t$ at step $t$ with the highest probability from the distribution to be injected into $d$.

\item We get the final perturbation $\mathcal{P} = \{p^1, p^2, \ldots, p^t, \ldots, p^T\}$, where $ T < \left| \mathbb{P} \right|$ and  $T$ is the number of steps.
\end{itemize}

\vspace{-2mm}
\subsection{Training with policy gradient} \label{training}
We solve the \ac{MDP} problem with the policy gradient algorithm REINFORCE \cite{sutton2018reinforcement}. 
In each episode, a trajectory $\tau = d^1,p^1,\ldots,d^T,p^T$ is sampled using policy $\pi$.
The episode terminates at step $T$, when term manipulation budget $\epsilon$ is reached, i.e. $\sum_{t=1}^{T} \left| p^t \right| \leq \epsilon < \sum_{t=1}^{T+1} \left| p^t \right|$.

The aim of training is to learn an optimal policy $\pi^*$ by maximizing the expected cumulative reward $R(\tau) = \mathbb{E}\left[\sum_{t=1}^{T} \gamma^t r^t \right]$,
where $\gamma \in [0,1)$ is the discount factor for future rewards. The training objective is to maximize $J(\phi,\varphi)$ via:
\begin{equation}
\nabla_{\phi,\varphi} J(\phi,\varphi) = \mathbb{E}_{\pi_{\phi,\varphi}}\left[\nabla_{\phi,\varphi} \log \pi_{\phi,\varphi} R(\tau)\right],
\end{equation}
where $\phi$ and $\varphi$ denote the parameter for sub-agent $I_{\phi}$ and meta-agent $G_{\varphi}$, respectively.
In \ac{RL-MARA}, the policy networks of two agents are spliced together to be optimized together.

The solution can be approximated by a Monte Carlo estimator \cite{LeventeKocsis2006BanditBM}, i.e., 
$
\nabla_{\phi,\varphi} J(\phi,\varphi) \propto \sum_{u=1}^{U}\sum_{t=1}^{T} \nabla_{\phi,\varphi} \log \pi_{\phi,\varphi}\left(p^{u,t}\mid d^{u,t}\right) R^{u,t},
$ where $U$ is the number of samples and $T$ is the number of steps.

\vspace{-2mm}
\subsection{Discussion}
It is important to point out two key differences between \ac{RL-MARA} and \ac{MoE} frameworks in \ac{IR} \cite{cai2023came, zou2022automatic}, which seem more straightforward in assembling existing ranking attack methods.
For MoE methods: 
\begin{enumerate*}[label=(\roman*)]
    \item The \ac{MoE} requires optimization of each expert (attacker) during training. 
    However, mainstream attack methods primarily use a series of search algorithms with two detached steps, finding vulnerable positions and adding perturbations, both of which are \textit{non-trainable}. 
    While it is feasible to design an algorithm for the first step, the second step poses a challenge, as generating discrete textual perturbations makes it difficult to obtain direct supervised signals. 
    \item Constrained by the gating, the outcome of \ac{MoE} frameworks is determined by one of a single expert and a weighting of all experts, introducing a sense of  \textit{rigidity}.  
\end{enumerate*} 

In contrast, for \ac{RL-MARA}: 
\begin{enumerate*}[label=(\roman*)]
\item \ac{RL-MARA} integrates existing single-granular attack methods in a \textit{trainable} manner. 
In addressing the two non-trainable steps present in existing attacks (see Section \ref{Sec:Pre}): 
to identify vulnerable positions, \ac{RL-MARA} employs a sub-agent called vulnerability indicator (see Section~\ref{Sec: Sub-agent}); 
to perturb the discrete text space, \ac{RL-MARA} uses a meta-agent called perturbation aggregator to master the perturbation addition strategy (see Section~\ref{Sec: meta-agent}); and 
\item  \ac{RL-MARA} chooses a perturbation of any granularity at each step. 
Consequently, by the end of the attack, we can achieve either a single-granular perturbation or a combination of different levels of perturbation granularity with \textit{flexibility}. 
\end{enumerate*}
\vspace*{-3mm}
\section{Experimental Settings}
\subsection{Datasets}
\textbf{Benchmark datasets.}
Like previous work \cite{wu2022prada,liu2023topic}, we conduct experiments on two benchmark datasets: 
\begin{enumerate*}[label=(\roman*)]
    \item The \textbf{MS MARCO Document Ranking dataset} \cite{nguyen2016ms} (MS MARCO) is a large-scale benchmark dataset for Web document retrieval, with about 3.21 million documents. 
    \item The \textbf{ClueWeb09-B dataset} \cite{clarke2009overview} (ClueWeb09) comprises 150 queries with a collection of 50 million documents, with 242 additional queries from the TREC Web Track 2012 \cite{clarke2012overview}.
\end{enumerate*}

\heading{Target queries and documents}
Following \cite{chen2023towards,liu2023topic}, we randomly sample 1000 Dev queries from MS MARCO and use 242 additional queries from ClueWeb09 as target queries for each dataset evaluation, respectively.
For each target query, we adopt two categories of target documents based on the top-100 ranked results of the target \ac{NRM}, considering different levels of attack difficulty, i.e., \textit{Easy} and \textit{Hard}.
Specifically, we randomly choose 5 documents ranked between $[30,60]$ as Easy target documents and select the 5 bottom-ranked documents as Hard target documents.
In addition to the two types, we incorporate \textit{Mixture} target documents for a thorough analysis. These consist of 5 documents randomly sampled from both the Easy and Hard target document sets.

\vspace{-2mm}
\subsection{Evaluation metrics}
\textbf{Attack performance.} 
We use three automatic metrics:
\begin{enumerate*}[label=(\roman*)]
\item Attack success rate (ASR) (\%), which evaluates the percentage of target documents successfully boosted under the corresponding target query. 
\item Average boosted ranks (Boost), which evaluates the average improved rankings for each target document under the corresponding target query.
\item Boosted top-$K$ rate (T$K$R) (\%), which evaluates the percentage of target documents that are boosted into top-$K$ under the corresponding target query.
\end{enumerate*}
The effectiveness of an adversary is better with a higher value for all these metrics. 

\heading{Naturalness performance}
Here, we use four metrics:
\begin{enumerate*}[label=(\roman*)]
\item Spamicity detection, which detects whether target web pages are spam.  
Following \cite{wu2022prada,liu2022order}, we adopt a utility-based term spamicity detection method, OSD \cite{zhou2009osd}, to detect the after-attack documents. 
\item Grammar checking, which calculates the average number of errors in the after-attack documents. 
Specifically, we use, Grammarly~\citep{grammarly}, an online grammar checker following~\cite{liu2022order}. 
\item Language model perplexity (PPL), which measures the fluency using the average perplexity calculated using a pre-trained GPT-2 model~\cite{radford2019language}.
\item Human evaluation, which measures the quality of the after-attack documents following the criteria in \cite{wu2022prada,liu2022order}. 
\end{enumerate*}

\vspace{-2mm}
\subsection{Models}
\textbf{Target NRMs.} 
We choose three typical \acp{NRM} as target \acp{NRM}:
\begin{enumerate*}[label=(\roman*)]
\item a pre-trained model, \emph{BERT} \cite{devlin2018bert}; 
\item a pre-trained model tailored for IR,  \emph{PROP} \cite{ma2021prop}; and 
\item a model distilled from the ranking capability of LLMs, \emph{RankLLM} \cite{sun-etal-2023-chatgpt}. 
\end{enumerate*}
For RankLLM, we employ the model introduced by \cite{sun-etal-2023-chatgpt}, distilling the ranking capabilities of an \ac{LLM}, i.e., ChatGPT, into DeBERTa-large \cite{he2020deberta} in a permutation distillation manner on LLM-generated permutations within MS MARCO.

\heading{Baselines} 
We compare the following attack methods against \acp{NRM}:  
\begin{enumerate*}[label=(\roman*)]
    \item \textbf{Term spamming (TS)}~\cite{gyongyi2005web} randomly selects a starting position in the target document and replaces the subsequent words with terms randomly sampled from the target query. 
    \item \textbf{PRADA}~\cite{wu2022prada}, \textbf{PLAT}~\cite{lei2022phrase} and \textbf{PAT}~\cite{liu2022order} are representative word-level, phrase-level and sentence-level ranking attack method against \acp{NRM}, introduced in Section \ref{Sec:Pre}.   
    \item \textbf{IDEM}~\cite{chen2023towards} is sentence-level ranking attack that inserts the generated connection sentence in the document. 
\end{enumerate*} 

\heading{Model variants} 
We implement three variants of \ac{RL-MARA}: 
\begin{enumerate*}[label=(\roman*)]
    \item \textbf{RL-}\allowbreak\textbf{MARA$_\mathrm{single}$} selects a single level of perturbation for each document. After generating a vulnerability distribution, it sums confidence scores for each perturbation level. The one with the highest total score is chosen, while others are ignored and their scores are reset to zero. 
    \item \textbf{RL-MARA$_\mathrm{triple}$} incorporates all three perturbation levels in each document. It calculates average confidence scores for each level across vulnerable spans, ranks them from high to low, and then sequentially selects the highest-ranked span at each level until the term manipulation budget is reached, ensuring each perturbation level occurs at least once.
    \item \textbf{RL-MARA$_\mathrm{greedy}$} greedily chooses vulnerable spans with the highest confidence for perturbation. It computes and ranks average confidence scores for vulnerable spans, regardless of granularity, and applies perturbations in descending order until the term manipulation budget is exhausted.
\end{enumerate*}

\begin{table*}[t]
\centering
   \caption{Attack performance of RL-MARA and baselines; $\ast$ indicates significant improvements over the best baseline ($p \le 0.05$).}
   \renewcommand{\arraystretch}{0.88}
   \setlength\tabcolsep{4.5pt}
  	\begin{tabular}{l  c c c c  c c c c   c c c c  c c c c }
  \toprule
   & \multicolumn{8}{c}{MS MARCO} & \multicolumn{8}{c}{ClueWeb09}  \\ 
  \cmidrule(r){2-9} \cmidrule(r){10-17} 
  Method & \multicolumn{4}{c}{Easy} & \multicolumn{4}{c}{Hard} & \multicolumn{4}{c}{Easy}& \multicolumn{4}{c}{Hard} \\
  \cmidrule{1-1} \cmidrule(r){2-5} \cmidrule(r){6-9} \cmidrule(r){10-13} \cmidrule(r){14-17}
   \textbf{BERT}    & ASR & Boost & T10R & T5R & ASR & Boost & T10R & T5R & ASR & Boost & T10R & T5R & ASR & Boost & T10R & T5R \\ 
       \midrule
TS & 100.0 & 38.1 & 84.3 & 26.9 & 89.5 & 68.2 & 23.6 & 5.9 
& 100.0 & 36.2 & 81.0 & 23.6 & 90.5 & 65.9 & 21.8 & 4.6\\
PRADA  & \phantom{1}98.3 & 26.1 & 69.3 & 18.3 & 78.9 & 55.9 & \phantom{1}9.6 & 1.8 
& \phantom{1}97.6 & 24.8 & 66.9 & 16.5 & 77.1 & 53.9 & \phantom{1}8.2 & 1.2\\
PLAT   & \phantom{1}93.6 & 24.3 & 63.1 & 15.6 & 72.1 & 50.0 & \phantom{1}8.5 & 1.1
& \phantom{1}92.1 & 23.1 & 61.9 & 14.0 & 70.2 & 48.2 & \phantom{1}7.2 & 0.8 \\
PAT   & 100.0 & 35.1 & 78.1 & 23.8 & 82.3 & 60.3 & 18.3 & 3.9 
& 100.0 & 34.3 & 75.6 & 20.6 & 78.3 & 54.1 & 14.9 & 2.1 \\
IDEM   & 100.0 & 39.6 & 85.6 & 26.8 & 90.2 & 69.6 & 25.8 & 7.2 
& 100.0 & 37.1 & 82.6 & 24.9 & 87.2 & 65.2 & 22.1 & 5.1\\
\midrule
RL-MARA$_\mathrm{single}$  & 100.0 & 36.4 & 82.6 & 26.0 & 88.7 & 67.2 & 23.9 & 6.1 
& 100.0 & 35.4 & 81.0 & 24.3 & 86.3 & 65.6 & 21.8 & 5.3\\
RL-MARA$_\mathrm{triple}$  & 100.0 & 40.3 & 87.1 & 27.9 & 93.5 & 75.1\rlap{$^{\ast}$} & 30.6\rlap{$^{\ast}$} & 8.6\rlap{$^{\ast}$}
& 100.0 & 38.9 & 85.7\rlap{$^{\ast}$} & 26.8 & 91.7\rlap{$^{\ast}$} & 73.2\rlap{$^{\ast}$} & 28.9\rlap{$^{\ast}$} & 7.8\rlap{$^{\ast}$}\\
RL-MARA$_\mathrm{greedy}$  & 100.0 & 40.8 & 88.3\rlap{$^{\ast}$} & 28.3\rlap{$^{\ast}$} & 94.6\rlap{$^{\ast}$} & 78.3\rlap{$^{\ast}$} & 31.3\rlap{$^{\ast}$} & 9.3\rlap{$^{\ast}$} 
& 100.0 & 38.8 & 86.1\rlap{$^{\ast}$} & 26.9\rlap{$^{\ast}$} & 93.7\rlap{$^{\ast}$} & 74.1\rlap{$^{\ast}$} & 29.2\rlap{$^{\ast}$} & 8.4\rlap{$^{\ast}$}\\
RL-MARA  & 100.0 & \textbf{42.2}\rlap{$^{\ast}$} & \textbf{90.2}\rlap{$^{\ast}$} & \textbf{30.2}\rlap{$^{\ast}$} & \textbf{98.9}\rlap{$^{\ast}$} & \textbf{88.1}\rlap{$^{\ast}$} & \textbf{36.9}\rlap{$^{\ast}$} & \textbf{9.9}\rlap{$^{\ast}$} 
& 100.0 & \textbf{40.1}\rlap{$^{\ast}$} & \textbf{87.9}\rlap{$^{\ast}$} & \textbf{28.6}\rlap{$^{\ast}$} & \textbf{97.2}\rlap{$^{\ast}$} & \textbf{85.8}\rlap{$^{\ast}$} & \textbf{35.3}\rlap{$^{\ast}$} & \textbf{9.5}\rlap{$^{\ast}$}\\
\midrule
   \textbf{PROP}  & ASR & Boost & T10R & T5R & ASR & Boost & T10R & T5R & ASR & Boost & T10R & T5R & ASR & Boost & T10R & T5R \\ 
       \midrule
TS & 100.0 & 37.6 & 83.0 & 25.8 & 89.7 & 67.3 & 22.8 & 5.1 
& 100.0 & 35.0 & 79.6 & 22.8 & 90.5 & 64.8 & 20.9 & 4.3\\
PRADA  & \phantom{1}95.2 & 23.4 & 66.6 & 16.4 & 75.8 & 53.4 & \phantom{1}8.6 & 1.2 
& \phantom{1}93.4 & 22.5 & 63.4 & 14.1 & 74.9 & 51.2 & \phantom{1}6.8 & 0.9\\
PLAT   & \phantom{1}91.2 & 22.0 & 60.5 & 13.5 & 69.9 & 48.2 & \phantom{1}7.5 & 0.8
& \phantom{1}89.8 & 21.4 & 59.8 & 12.1 & 68.0 & 46.3 & \phantom{1}6.7 & 0.5 \\
PAT   & \phantom{1}98.6 & 33.6 & 75.9 & 22.5 & 80.2 & 58.7 & 17.3 & 3.1 
& \phantom{1}96.3 & 31.1 & 72.2 & 20.1 & 77.3 & 53.9 & 15.1 & 2.4 \\
IDEM   & 100.0 & 37.3 & 83.0 & 25.0 & 87.9 & 67.5 & 24.0 & 6.5 
& 100.0 & 35.8 & 80.1 & 23.1 & 85.8 & 65.1 & 21.7 & 4.1\\
\midrule
RL-MARA$_\mathrm{single}$  & 100.0 & 35.3 & 80.2 & 24.5 & 86.4 & 65.1 & 23.0 & 5.8 
& 100.0 & 34.6 & 79.9 & 23.1 & 84.5 & 63.5 & 20.0 & 4.9\\
RL-MARA$_\mathrm{triple}$  & 100.0 & 37.9 & 85.6 & 26.5 & 91.2 & 73.8\rlap{$^{\ast}$} & 28.9\rlap{$^{\ast}$} & 8.0\rlap{$^{\ast}$}
& 100.0 & 37.5 & 83.8\rlap{$^{\ast}$} & 25.3\rlap{$^{\ast}$} & 90.1\rlap{$^{\ast}$} & 71.8\rlap{$^{\ast}$} & 27.2\rlap{$^{\ast}$} & 6.3\rlap{$^{\ast}$}\\
RL-MARA$_\mathrm{greedy}$  & 100.0 & 38.4 & 87.9\rlap{$^{\ast}$} & 27.6\rlap{$^{\ast}$} & 92.3\rlap{$^{\ast}$} & 77.6\rlap{$^{\ast}$} & 30.8\rlap{$^{\ast}$} & 8.7\rlap{$^{\ast}$} 
& 100.0 & 36.3 & 85.1\rlap{$^{\ast}$} & 25.4\rlap{$^{\ast}$} & 91.9\rlap{$^{\ast}$} & 72.5\rlap{$^{\ast}$} & 28.5\rlap{$^{\ast}$} & 7.9\rlap{$^{\ast}$}\\
RL-MARA  & 100.0 & \textbf{41.0}\rlap{$^{\ast}$} & \textbf{88.7}\rlap{$^{\ast}$} & \textbf{28.9}\rlap{$^{\ast}$} & \textbf{97.5}\rlap{$^{\ast}$} & \textbf{87.0}\rlap{$^{\ast}$} & \textbf{36.0}\rlap{$^{\ast}$} & \textbf{9.1}\rlap{$^{\ast}$} 
& 100.0 & \textbf{39.2}\rlap{$^{\ast}$} & \textbf{85.6}\rlap{$^{\ast}$} & \textbf{27.4}\rlap{$^{\ast}$} & \textbf{95.8}\rlap{$^{\ast}$} & \textbf{83.6}\rlap{$^{\ast}$} & \textbf{33.8}\rlap{$^{\ast}$} & \textbf{8.9}\rlap{$^{\ast}$}\\
\midrule
   \textbf{RankLLM} & ASR & Boost & T10R & T5R & ASR & Boost & T10R & T5R & ASR & Boost & T10R & T5R & ASR & Boost & T10R & T5R \\ 
       \midrule
TS & 100.0 & 34.3 & 79.4 & 22.5 & 89.8 & 63.9 & 19.7 & 3.2 
& \phantom{1}99.2 & 32.1 & 73.0 & 19.3 & 89.9 & 59.8 & 18.4 & 2.8\\
PRADA  & \phantom{1}92.1 & 21.1 & 60.9 & 13.4 & 68.9 & 50.2 & \phantom{1}6.7 & 0.7 
& \phantom{1}89.6 & 20.2 & 59.8 & 12.3 & 70.3 & 48.9 & \phantom{1}5.2 & 0.5\\
PLAT   & \phantom{1}88.9 & 19.1 & 55.8 & 11.5 & 63.5 & 45.9 & \phantom{1}5.9 & 0.5
& \phantom{1}85.6 & 18.6 & 55.9 & 10.0 & 64.8 & 42.6 & \phantom{1}5.2 & 0.3 \\
PAT   & \phantom{1}95.6 & 30.2 & 72.1 & 19.8 & 75.6 & 54.3 & 14.9 & 2.8 
& \phantom{1}93.9 & 28.7 & 68.6 & 18.4 & 73.5 & 49.8 & 12.8 & 1.7 \\
IDEM   & \phantom{1}98.9 & 34.8 & 79.2 & 22.2 & 84.8 & 63.2 & 21.8 & 5.2 
& \phantom{1}98.2 & 33.2 & 77.9 & 21.1 & 82.3 & 61.8 & 18.9 & 3.2\\
\midrule
RL-MARA$_\mathrm{single}$  & \phantom{1}97.9 & 33.6 & 77.9 & 22.3 & 83.8 & 62.6 & 21.2 & 4.3 
& \phantom{1}97.5 & 32.1 & 76.7 & 21.2 & 81.8 & 60.2 & 18.1 & 3.7\\
RL-MARA$_\mathrm{triple}$  & \phantom{1}99.8 & 35.8 & 82.9 & 24.3 & 89.0 & 71.8\rlap{$^{\ast}$} & 27.0\rlap{$^{\ast}$} & 7.1\rlap{$^{\ast}$}
& \phantom{1}99.2 & 35.2 & 81.2\rlap{$^{\ast}$} & 22.8\rlap{$^{\ast}$} & 88.7\rlap{$^{\ast}$} & 69.8\rlap{$^{\ast}$} & 25.4\rlap{$^{\ast}$} & 5.8\rlap{$^{\ast}$}\\
RL-MARA$_\mathrm{greedy}$  & 100.0 & 36.2\rlap{$^{\ast}$} & 85.3\rlap{$^{\ast}$} & 25.1\rlap{$^{\ast}$} & 89.7\rlap{$^{\ast}$} & 74.8\rlap{$^{\ast}$} & 28.9\rlap{$^{\ast}$} & 8.1\rlap{$^{\ast}$} 
& \phantom{1}99.7 & 34.6 & 82.3\rlap{$^{\ast}$} & 22.7\rlap{$^{\ast}$} & 89.2\rlap{$^{\ast}$} & 70.1\rlap{$^{\ast}$} & 26.4\rlap{$^{\ast}$} & 7.2\rlap{$^{\ast}$}\\
RL-MARA  & 100.0 & \textbf{39.7}\rlap{$^{\ast}$} & \textbf{85.8}\rlap{$^{\ast}$} & \textbf{27.0}\rlap{$^{\ast}$} & \textbf{95.6}\rlap{$^{\ast}$} & \textbf{85.0}\rlap{$^{\ast}$} & \textbf{34.3}\rlap{$^{\ast}$} & \textbf{8.6}\rlap{$^{\ast}$} 
& \textbf{100.0} & \textbf{37.0}\rlap{$^{\ast}$} & \textbf{82.4}\rlap{$^{\ast}$} & \textbf{25.2}\rlap{$^{\ast}$} & \textbf{92.1}\rlap{$^{\ast}$} & \textbf{81.1}\rlap{$^{\ast}$} & \textbf{31.3}\rlap{$^{\ast}$} & \textbf{8.2}\rlap{$^{\ast}$}\\
\bottomrule
    \end{tabular}
   \label{table:Baseline}
   \vspace{-5mm}
\end{table*}

\vspace{-2mm}
\subsection{Implementation details} \label{sec: details}
For MS MARCO and ClueWeb09, following \cite{wu2022prada}, we truncate each document to 512.
The initial retrieval is performed using the Anserini toolkit \cite{PeilinYang2018AnseriniRR} with the BM25 model to obtain the top 100 ranked documents following \cite{wu2022prada, liu2023topic}.
For the environment, following \cite{wu2022prada, liu2022order}, we use BERT$_{base}$ as the surrogate model, and the training details are consistent with \cite{wu2022prada, liu2023topic}. 
For the reward, the balance hyper-parameter $\beta$ is 0.2 and the discount factor $\gamma$ is 0.9.

For the perturbations, we set the term manipulation budget $\epsilon=25$ for RL-MARA and all baselines. 
We specify a word length of 1 for word-level perturbations, 2 to 5 for phrase-level perturbations, and 6 to 10 for sentence-level perturbations.
For the sub-agent, we supervise its training in a sequence labeling manner \cite{tsai2019small} to ensure that the labeling of each perturbation position is a continuous span.
We set the learning rate to $3e^{-6}$ with Adam as the optimizer to train \ac{RL-MARA}.
Following \cite{liu2023topic}, when the training process ends, we stop the updating of policy networks while running another epoch on the full dataset as a testing phase to evaluate the performance. 

\vspace{-2mm}
\section{Experimental Results}
\subsection{Attack evaluation} \label{Sec: Attack evaluation}
Table~\ref{table:Baseline} showcases the attack performance among three target \acp{NRM} with different attack methods, evaluated on both Easy and Hard target documents.
We have the following observations: 
\begin{enumerate*}[label=(\roman*)]
    \item Overall, the attack methods have effects on all three \acp{NRM}, exposing the prevalence of adversarial vulnerability.
    RankLLM is more resistant to adversarial attacks than other \acp{NRM}, indicating that distilling the ranking capabilities with \acp{LLM} helps to enhance the adversarial robustness of \acp{NRM}.
    \item The attack efficacy of most methods on ClueWeb09 is observed to be inferior compared to their performance on MS MARCO.
    This disparity may stem from the noise present in ClueWeb09's documents, which potentially renders the model less responsive to adversarial perturbations. 
    This observation aligns with previous findings reported in \cite{liu2023topic}.
    \item Hard documents exhibit lower ASR and T$K$R compared to Easy ones, due to the higher prevalence of irrelevant information in bottom-ranked documents, which challenges effective attacks with limited perturbation.
    \item Sentence-level attack methods yield better attack results than word-level and phrase-level methods.
    The reason may be that the sentence-level perturbation is a continuous optimization of an entire vulnerable span in a document, thus increasing the likelihood of misleading the relevance judgment of \acp{NRM} to a greater extent.
    However, these sentence-level attack methods run the risk of being suspect due to naturalness flaws in respective aspects.
    We will discuss this further in Section \ref{Naturalness}.
\end{enumerate*}

\ac{RL-MARA} significantly outperforms all baselines. In Hard documents of MS MARCO, while attacking RankLLM, \ac{RL-MARA} improves over the best baseline, IDEM, by 65.4\% in T5R and by 34.5\% in Boost, highlighting the attack effect of perturbations at multiple levels of granularity. 
\begin{enumerate*}[label=(\roman*)]
\item The superiority of \ac{RL-MARA} over RL-MARA$_\mathrm{single}$ suggests that the single granularity of perturbation is not sufficient to fully utilize the diverse vulnerability distributions in documents.
Moreover, the synergy between perturbations of different levels of granularity can lead to more threatening adversarial examples.
\item The advantage of \ac{RL-MARA} over RL-MARA$_\mathrm{triple}$ indicates that 
the sequential decision-making in adding perturbations of each granularity, based on \ac{RL} rather than mechanically applying perturbations across all levels of granularity, enables more effective exploitation of each document's vulnerability distribution.
\item The improvement of \ac{RL-MARA} over RL-MARA$_\mathrm{greedy}$ demonstrates that the cooperative approach of two agents in flexibly organizing multi-granular perturbations is instrumental in generating high-quality perturbation sequences, posing significant threats to \acp{NRM}.
\end{enumerate*}

For the Mixture target documents, the performance of all attack methods remains consistent with that observed in each \ac{NRM} for Easy and Hard documents.
Even attacking the most defensive Rank\-LLM on MS MARCO, \ac{RL-MARA} outperforms the best baseline by 34.8\% in T5R and 25.1\% in Boost. 

\begin{table}[t]
\centering
   \caption{The MRR@10 (\%) performance on MS MARCO and ClueWeb09 of target NRMs (BERT, PROP and RankLLM) and their corresponding surrogate models (S$_\mathrm{NRM}$). }
   \renewcommand{\arraystretch}{0.92}
   \setlength\tabcolsep{2.3pt}
  	\begin{tabular}{l c c  c c   c c}
  \toprule
       NRM & BERT & S$_\mathrm{BERT}$ & RROP & S$_\mathrm{RROP}$ & RankLLM & S$_\mathrm{RankLLM}$ \\
       \midrule
MS MARCO & 38.48 & 35.41 & 39.01 & 36.24 & 39.89 & 37.86 \\
ClueWeb09 & 27.50 & 24.93 & 28.25 & 25.46 & 28.96 & 26.65\\
\bottomrule
    \end{tabular}
   \label{table:ranking performance}
   \vspace*{-1mm}
\end{table}

Furthermore, the performance of the surrogate ranking model plays an important role in the success of the black-box attack.
As shown in Table \ref{table:ranking performance}, for all target NRMs, the corresponding surrogate model can imitate their performance to some extent. 
This allows the vulnerabilities identified on the surrogate ranking model to be effectively transferred to the target \ac{NRM}.

\begin{figure}[t]
    \centering
    \includegraphics[width=\linewidth]{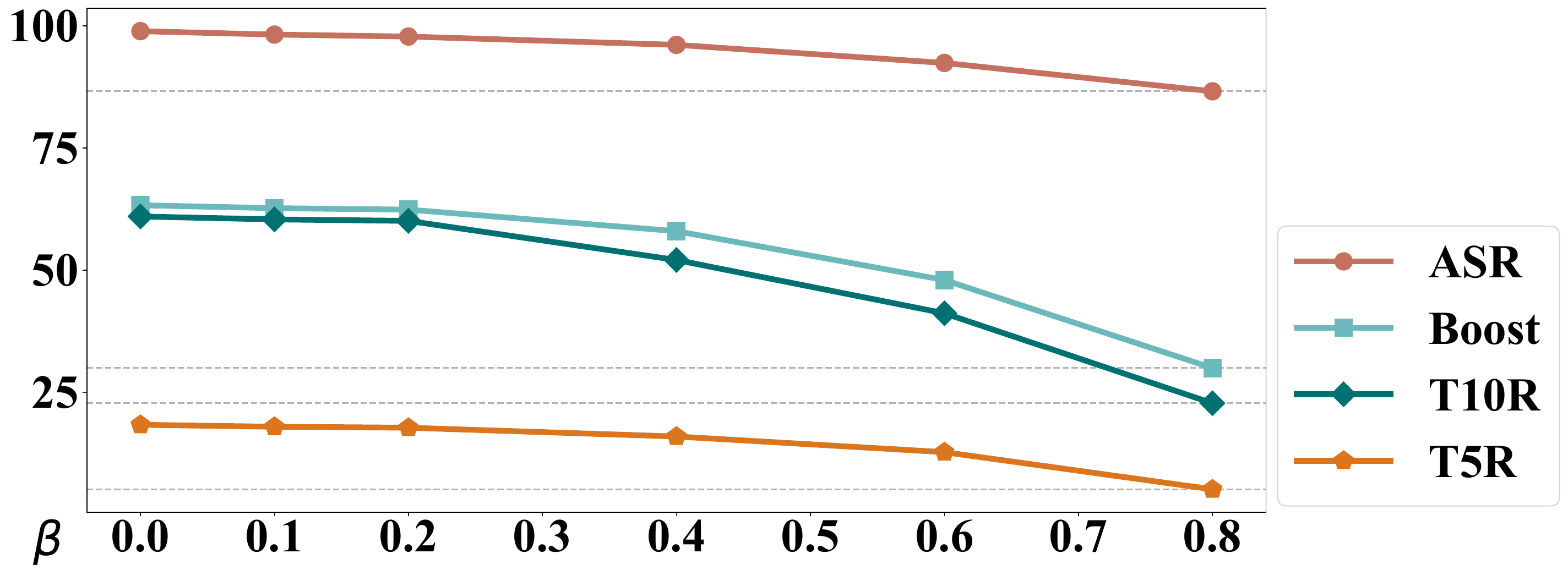}
    \caption{The impact of hyper-parameter $\beta$ on the attack performance of RL-MARA against RankLLM on MS MARCO.}
    \label{fig:both}
    \vspace*{-2mm}
\end{figure}
\vspace*{-1mm}

\heading{The impact of the balance hyper-parameter}
$\beta$ is an important hyper-parameter in our multi-granular reward, since it balances the attack performance and the naturalness of adversarial example.
We investigate the impact of $\beta$ on the performance of \ac{RL-MARA}.
Lower values of $\beta$ imply a less emphasis on the naturalness and a greater emphasis on the attack effectiveness.
We take attacking RankLLM on Mixture documents of MS MARCO as an example.
The trend of the attack performance with $\beta$ is shown in Figure \ref{fig:both}.
As $\beta$ decreases, the attack performance gradually increases and plateaus, since the naturalness reward no longer dominates the update of the attack strategy.
However, trivializing naturalness reward signals can lead to susceptibility to suspicion, as discussed in Section~\ref{Naturalness}.

When serving as a benchmark, \ac{RL-MARA} can create adversarial examples of varying naturalness by tuning the hyper-parameter $\beta$, enabling a comprehensive analysis for model robustness.

\vspace{-2mm}
\subsection{Capability of surrogate models}
\textbf{Black-box vs. white-box attack.}
We further focus on the white-box scenario, which is valuable for enhancing understanding of our method.
In white-box attacks, we directly substitute the surrogate ranking model as the target \ac{NRM} and keep other components the same in \ac{RL-MARA}.
We consider the most defensive RankLLM, which has a 4.4\% higher ranking performance than its surrogate model, as shown in Table~\ref{table:ranking performance}.
The result on the Mixture target documents of MS MARCO is shown in Figure~\ref{fig:both} (Left),  with similar findings on other target documents.
Compared with the white-box setting, \ac{RL-MARA} still obtains competitive performance in black-box scenarios.
The results demonstrate that the training method of the surrogate model is sufficient to simulate the vulnerability performance of the target \ac{NRM} at different levels of granularity, thus making our multi-granular attack effects transferable.

\heading{Training a surrogate model in the out-of-distribution (OOD) setting}
In our experiments, the training data of the surrogate model is directly adopted from the Eval queries of the target model, i.e., the same distribution as the query used to train the target \ac{NRM}.
However, in realistic search scenarios, obtaining an identically distributed query set is difficult. 
Following \cite{chen2023towards}, we show the results for the IID and OOD  scenarios in Figure \ref{fig:both} (Right). 
Specifically, we take the Mixture target documents as an example and evaluate the attack performance against RankLLM on MS MARCO. 
For our IID scenarios, we use Eval queries of MS MARCO for surrogate model training. 
For OOD scenarios, we use Eval queries of Natural Questions (NQ) \cite{kwiatkowski2019natural} to train the surrogate model and observe a 25.7\% decrease on MRR@10 relative to the IID surrogate model. 
The results reveal that despite compromised attack performance when the IID data is unavailable, \ac{RL-MARA} continues to perform an effective attack method that can identify model vulnerabilities.

\begin{figure}[t]
    \centering
    \includegraphics[width=\linewidth]{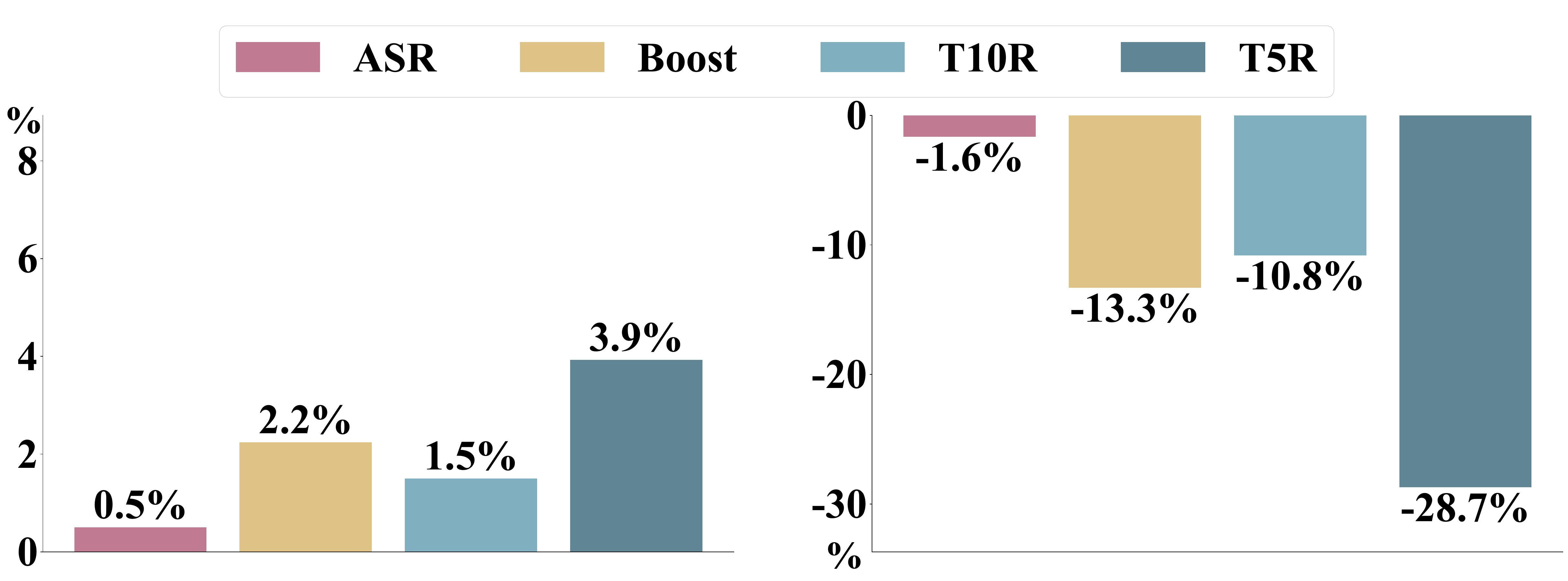}
    \caption{(Left): Attack performance changes of RL-MARA against RankLLM on MS MARCO in the white-box setting, compared to black-box setting. (Right): Attack performance changes of RL-MARA against RankLLM on MS MARCO in the OOD scenario, compared to the IID scenario.}
    \label{fig:both}
    \vspace*{-2mm}
\end{figure}
\vspace*{-2mm}

\begin{table}[t]
\centering
   \caption{The online grammar checker, perplexity, and human evaluation results for attacking RankLLM on MS MARCO.}
   \renewcommand{\arraystretch}{0.9}
   \setlength\tabcolsep{1pt}
  	\begin{tabular}{l c c  c c   c c}
  \toprule
       Method & Grammar & PPL & Impercept. & \textit{kappa} & Fluency & \textit{Kendall} \\
       \midrule
Original & 59 & 44.2 & 0.90 & 0.52 & 4.57 & 0.65\\
\midrule
TS  & 67 & 59.7 & 0.06 & 0.53 & 2.42 & 0.73\\
PRADA & 108 & 113.4 & 0.53 & 0.48 & 3.21 & 0.82\\
PLAT  & \phantom{1}98 & \phantom{1}86.3 & 0.62 & 0.58 & 3.30 & 0.74 \\
PAT & \phantom{1}83 & \phantom{1}72.1 & 0.58 & 0.61 & 3.42 & 0.76\\
IDEM & \phantom{1}67 & \phantom{1}55.6 & 0.79 & 0.46 & 3.75 & 0.82 \\
RL-MARA$_\mathrm{-NC}$ & 103 & \phantom{1}89.5 & 0.58 & 0.47 & 2.97 & 0.69 \\
RL-MARA & \phantom{1}65 & \phantom{1}53.5 & 0.87 & 0.58 & 3.90 & 0.76\\
\bottomrule
    \end{tabular}
   \label{table:human evaluation}
   \vspace{-3mm}
\end{table}

\vspace{-2mm}
\subsection{Naturalness evaluation}\label{Naturalness}
Next, we report on the naturalness of generated adversarial examples against RankLLM on MS MARCO, with similar findings on ClueWeb09.
Here, we set the balance hyper-parameter $\beta$ to $0$ of \ac{RL-MARA}, denoted as RL-MARA$_\mathrm{-NC}$, for comparison.

\heading{Grammar checking, PPL, and human evaluation}
Table \ref{table:human evaluation} lists the results of the automatic grammar checker, PPL, and human
evaluation.
For human evaluation, we recruit five annotators to annotate 32 randomly sampled Mixture adversarial examples from each attack method \cite{chen2023towards}; 
The annotators score the \emph{Fluency} from 1 to 5; higher scores indicate more fluent examples; the \emph{Imperceptibility} is to determine whether an example is attacked (0) or not (1); including the annotation consistency (the \textit{Kappa} value and \textit{Kendall's Tau} coefficient)  \cite{wu2022prada, liu2023topic}. 
We observe that: 
\begin{enumerate*}[label=(\roman*)]
\item TS demonstrates subpar performance in all naturalness evaluations because it abruptly inserts the query terms into the document without regard to semantic coherence.
\item The single-granular attack methods fall short when compared to the original documents, a possible reason is that \textit{a priori} restricting perturbations to a specific level of granularity carries the risk of generating unnatural perturbations. 
\item Although discarding the fluency constraint may improve attack efficacy, the adversarial examples produced by RL-MARA$_{-NC}$ tend to arouse suspicion.
\item \ac{RL-MARA} outperforms the baselines, demonstrating the effectiveness of the naturalness reward provided by the \ac{LLM}.
\end{enumerate*}

\begin{table}[t]
\centering
   \caption{The detection rate (\%) via a representative anti-spamming method for attacking RankLLM on MS MARCO.}
   \renewcommand{\arraystretch}{0.88}
   \setlength\tabcolsep{12pt}
  	\begin{tabular}{l c c c c}
  \toprule
       Threshold & 0.08 & 0.06 & 0.04 & 0.02 \\ 
       \midrule
TS  & 40.3 & 53.4 & 76.3 & 94.5 \\
PRADA  & 13.4 & 22.6 & 40.1 & 64.2 \\
PLAT   & 11.6 & 20.6 & 32.5 & 58.7 \\
PAT   & \phantom{1}9.3 & 15.2 & 26.4 & 49.8 \\
IDEM   & 17.2 & 29.4 & 48.3 & 72.9 \\
RL-MARA$_\mathrm{-NC}$ & 15.8 & 26.3 & 42.4 & 68.6 \\
RL-MARA & \phantom{1}\textbf{6.2} & \textbf{10.6} & \textbf{20.2} & \textbf{40.3} \\
\bottomrule
    \end{tabular}
   \label{table:anti-spamming}
\end{table}

\heading{Spamicity detection}
Table \ref{table:anti-spamming} shows the automatic spamicity detection results on Mixture documents with similar findings on other target documents.
If the spamicity score of a document exceeds the detection threshold, it is identified as suspected spam content.
Even with the effective attack performance, IDEM is easier to detect among other single-granular attack baselines because it ignores the avoidance of query terms in prompting the language model to generate perturbed sentences.
\ac{RL-MARA} outperforms the baselines significantly (p-value $\leq 0.05$), demonstrating the similarity reward provided by an \ac{LLM} is effective in preventing attack methods from abusing query terms and thus being suspected.

\vspace*{-2mm}
\section{Related Work}
\textbf{Neural ranking models.} 
Ranking models have evolved from early heuristic models \cite{GerardSalton1975AVS} to probabilistic \cite{StephenRobertson1994SomeSE,JayPonte1998ALM}, and modern learning-to-rank models \cite{li2014learning,TieYanLiu2009LearningTR}. 
\acp{NRM} \cite{ZhuyunDai2019DeeperTU,guo2016deep,KezbanDilekOnal2018NeuralIR} emerged with deep learning and demonstrated excellent ranking performance through powerful relevance modeling capabilities.
Research has also explored integrating potent pre-trained language models into ranking tasks, including tailoring specific pre-training \cite{ma2021prop}, fine-tuning \cite{RodrigoNogueira2019PassageRW}, and distilling techniques \cite{sun-etal-2023-chatgpt}, leading to state-of-the-art performance \cite{fan2022pre}. 
However, these NRMs also exhibit disconcerting adversarial vulnerabilities inherited from neural networks \cite{raval2020one,wang2022bert,liu2024perturbation,liu2023robustness}.

\heading{Adversarial ranking attacks}
With the development of deep learning, many fields such as retrieval augmentation \cite{zhang2023relevance}, recommender systems \cite{wang2021discover,wang2022micro}, and knowledge graphs \cite{liang2023message,Liangke_Survey} have begun to focus on the robustness challenge.
In the field of \ac{IR}, the challenge of black-hat search engine optimization (SEO) has been significant since the inception of (web) search engines~\cite{gyongyi2005web}. 
Typically, the goal of black-hat SEO is to boost a page's rank by maliciously manipulating documents in a way that is unethical or non-compliant with search engine guidelines~\cite{castillo2011adversarial,kurland2022competitive}.
It usually causes erosion of search quality and deterioration of the user experience.
To safeguard NRMs from exploitation, research has focused on adversarial ranking attacks \cite{wu2022prada,liu2022order}, in order to expose the vulnerability of NRMs.
The goal of adversarial ranking attacks is to manipulate the target document through imperceptible perturbations to improve its ranking for an individual or a small group of target queries \cite{wu2022prada,liu2023topic}. 
Previous studies have mainly explored attacks against NRMs with a single granularity of perturbations, e.g., word-level  \cite{wu2022prada,raval2020one,wang2022bert,liu2023topic,liu2023black} and sentence-level \cite{CongzhengSong2020AdversarialSC, liu2022order, liu2023topic, chen2023towards}. 
However, these efforts lack the flexibility needed for the comprehensive exploitation of NRM vulnerabilities. 
In this paper, we launch attacks with perturbations at multiple levels of granularity.

\heading{Multi-granular adversarial attack}
Multi-granular attacks, in contrast to single-granular ones, engage different levels of perturbation granularity \cite{zhang2020adversarial}.
In CV \cite{jin2023adversarial,wu2023imperceptible} and NLP \cite{chen2021multi, ebrahimi2018hotflip,li2019textbugger}, 
they exploit model vulnerabilities by freely combining varying perturbation granularities to produce threatening adversarial examples.
However, in NLP, existing multi-granular attack methods are mainly confined to selecting one granularity from several options in a predetermined manner \cite{chen2021multi,ebrahimi2018hotflip}.
Unlike these studies, we combine perturbations at multiple levels of granularity, allowing them to coexist in an adversarial example to enhance the effectiveness of adversarial ranking attacks.

\vspace*{-1mm}
\section{Conclusion}
In this work, we investigated a multi-granular ranking attack framework against black-box \acp{NRM}.
We modeled multi-granular attacks as a sequential decision-making process and proposed a reinforcement learning-based framework, \ac{RL-MARA}, to address it.
Our extensive experimental results reveal that the proposed method can effectively boost the target document through multi-granularity perturbations with imperceptibility.

In future work, we would like to explore how to efficiently optimize the combinatorial explosion problem due to multi-granular attacks from a theoretical perspective.
Beyond the current single-granular attack methods, we will explore integrating more granularity and different attack methods into our framework and making them learnable.
Our method proves effective against NRMs distilled from LLMs, as exemplified by RankLLM. However, directly attacking \acp{LLM} as rankers and, in turn, leveraging their advanced capabilities for attacks presents promising future research avenues.

\begin{acks}
This work was funded by the Strategic Priority Research Program of the CAS under Grants No. XDB0680102, 
the National Key Research and Development Program of China under Grants No. 2023YFA1011\\602 and 2021QY1701,  
the National Natural Science Foundation of China (NSFC) under Grants No. 62372431, 
the Youth Innovation Promotion Association CAS under Grants No. 2021100, 
the Lenovo-CAS Joint Lab Youth Scientist Project, 
and the project under Grants No. JCKY2022130C039.  
This work was also (partially) funded by 
the Hybrid Intelligence Center, a 10-year program funded by the Dutch Ministry of Education, Culture and Science through the Netherlands Organisation for Scientific Research, \url{https://hybrid-intelligence-centre.nl}, 
project LESSEN with project number NWA.1389.20.183 of the research program NWA ORC 2020/21, which is (partly) financed by the Dutch Research Council (NWO),
and
the FINDHR (Fairness and Intersectional Non-Discrimination in Human Recommendation) project that received funding from the European Union’s Horizon Europe research and innovation program under grant agreement No 101070212. 

All content represents the opinion of the authors,
which is not necessarily shared or endorsed by their respective employers and/or sponsors. 
\end{acks}

\bibliographystyle{ACM-Reference-Format}
\balance
\bibliography{references}


\begin{thebibliography}{71}


\ifx \showCODEN    \undefined \def \showCODEN     #1{\unskip}     \fi
\ifx \showDOI      \undefined \def \showDOI       #1{#1}\fi
\ifx \showISBNx    \undefined \def \showISBNx     #1{\unskip}     \fi
\ifx \showISBNxiii \undefined \def \showISBNxiii  #1{\unskip}     \fi
\ifx \showISSN     \undefined \def \showISSN      #1{\unskip}     \fi
\ifx \showLCCN     \undefined \def \showLCCN      #1{\unskip}     \fi
\ifx \shownote     \undefined \def \shownote      #1{#1}          \fi
\ifx \showarticletitle \undefined \def \showarticletitle #1{#1}   \fi
\ifx \showURL      \undefined \def \showURL       {\relax}        \fi
\providecommand\bibfield[2]{#2}
\providecommand\bibinfo[2]{#2}
\providecommand\natexlab[1]{#1}
\providecommand\showeprint[2][]{arXiv:#2}

\bibitem[Bellman(1957)]%
        {bellman1957markovian}
\bibfield{author}{\bibinfo{person}{Richard Bellman}.} \bibinfo{year}{1957}\natexlab{}.
\newblock \showarticletitle{A Markovian Decision Process}.
\newblock \bibinfo{journal}{\emph{Journal of mathematics and mechanics}} (\bibinfo{year}{1957}), \bibinfo{pages}{679--684}.
\newblock


\bibitem[Brendel et~al\mbox{.}(2018)]%
        {brendel2021decision}
\bibfield{author}{\bibinfo{person}{Wieland Brendel}, \bibinfo{person}{Jonas Rauber}, {and} \bibinfo{person}{Matthias Bethge}.} \bibinfo{year}{2018}\natexlab{}.
\newblock \showarticletitle{Decision-Based Adversarial Attacks: Reliable Attacks Against Black-Box Machine Learning Models}. In \bibinfo{booktitle}{\emph{ICLR}}.
\newblock


\bibitem[Cai et~al\mbox{.}(2023)]%
        {cai2023came}
\bibfield{author}{\bibinfo{person}{Yinqiong Cai}, \bibinfo{person}{Yixing Fan}, \bibinfo{person}{Keping Bi}, \bibinfo{person}{Jiafeng Guo}, \bibinfo{person}{Wei Chen}, \bibinfo{person}{Ruqing Zhang}, {and} \bibinfo{person}{Xueqi Cheng}.} \bibinfo{year}{2023}\natexlab{}.
\newblock \showarticletitle{CAME: Competitively Learning a Mixture-of-Experts Model for First-stage Retrieval}.
\newblock \bibinfo{journal}{\emph{arXiv preprint arXiv:2311.02834}} (\bibinfo{year}{2023}).
\newblock


\bibitem[Castillo and Davison(2011)]%
        {castillo2011adversarial}
\bibfield{author}{\bibinfo{person}{Carlos Castillo} {and} \bibinfo{person}{Brian~D. Davison}.} \bibinfo{year}{2011}\natexlab{}.
\newblock \showarticletitle{Adversarial Web Search}.
\newblock \bibinfo{journal}{\emph{Foundations and Trends in Information Retrieval}} \bibinfo{volume}{4}, \bibinfo{number}{5} (\bibinfo{year}{2011}), \bibinfo{pages}{377--486}.
\newblock


\bibitem[Chen et~al\mbox{.}(2023)]%
        {chen2023towards}
\bibfield{author}{\bibinfo{person}{Xuanang Chen}, \bibinfo{person}{Ben He}, \bibinfo{person}{Zheng Ye}, \bibinfo{person}{Le Sun}, {and} \bibinfo{person}{Yingfei Sun}.} \bibinfo{year}{2023}\natexlab{}.
\newblock \showarticletitle{Towards Imperceptible Document Manipulations against Neural Ranking Models}. In \bibinfo{booktitle}{\emph{ACL}}. \bibinfo{pages}{6648--6664}.
\newblock


\bibitem[Chen et~al\mbox{.}(2021)]%
        {chen2021multi}
\bibfield{author}{\bibinfo{person}{Yangyi Chen}, \bibinfo{person}{Jin Su}, {and} \bibinfo{person}{Wei Wei}.} \bibinfo{year}{2021}\natexlab{}.
\newblock \showarticletitle{Multi-granularity Textual Adversarial Attack with Behavior Cloning}. In \bibinfo{booktitle}{\emph{EMNLP}}. \bibinfo{pages}{4511--4526}.
\newblock


\bibitem[Clarke et~al\mbox{.}(2009)]%
        {clarke2009overview}
\bibfield{author}{\bibinfo{person}{Charles~L Clarke}, \bibinfo{person}{Nick Craswell}, {and} \bibinfo{person}{Ian Soboroff}.} \bibinfo{year}{2009}\natexlab{}.
\newblock \bibinfo{booktitle}{\emph{Overview of the TREC 2009 Web Track}}.
\newblock \bibinfo{type}{{T}echnical {R}eport}. \bibinfo{institution}{Waterloo University}.
\newblock


\bibitem[Clarke et~al\mbox{.}(2012)]%
        {clarke2012overview}
\bibfield{author}{\bibinfo{person}{Charles~L Clarke}, \bibinfo{person}{Nick Craswell}, {and} \bibinfo{person}{Ellen~M Voorhees}.} \bibinfo{year}{2012}\natexlab{}.
\newblock \bibinfo{booktitle}{\emph{Overview of the TREC 2012 Web Track}}.
\newblock \bibinfo{type}{{T}echnical {R}eport}. \bibinfo{institution}{NIST Gaithersburg MD}.
\newblock


\bibitem[Dai and Callan(2019)]%
        {ZhuyunDai2019DeeperTU}
\bibfield{author}{\bibinfo{person}{Zhuyun Dai} {and} \bibinfo{person}{Jamie Callan}.} \bibinfo{year}{2019}\natexlab{}.
\newblock \showarticletitle{Deeper Text Understanding for IR with Contextual Neural Language Modeling}. In \bibinfo{booktitle}{\emph{SIGIR}}.
\newblock


\bibitem[Dehghani et~al\mbox{.}(2017)]%
        {dehghani2017neural}
\bibfield{author}{\bibinfo{person}{Mostafa Dehghani}, \bibinfo{person}{Hamed Zamani}, \bibinfo{person}{Aliaksei Severyn}, \bibinfo{person}{Jaap Kamps}, {and} \bibinfo{person}{W~Bruce Croft}.} \bibinfo{year}{2017}\natexlab{}.
\newblock \showarticletitle{Neural Ranking Models with Weak Supervision}. In \bibinfo{booktitle}{\emph{SIGIR}}. \bibinfo{pages}{65--74}.
\newblock


\bibitem[Ebrahimi et~al\mbox{.}(2018)]%
        {ebrahimi2018hotflip}
\bibfield{author}{\bibinfo{person}{Javid Ebrahimi}, \bibinfo{person}{Anyi Rao}, \bibinfo{person}{Daniel Lowd}, {and} \bibinfo{person}{Dejing Dou}.} \bibinfo{year}{2018}\natexlab{}.
\newblock \showarticletitle{HotFlip: White-Box Adversarial Examples for Text Classification}. In \bibinfo{booktitle}{\emph{ACL}}. \bibinfo{pages}{31--36}.
\newblock


\bibitem[Fan et~al\mbox{.}(2018)]%
        {fan2018modeling}
\bibfield{author}{\bibinfo{person}{Yixing Fan}, \bibinfo{person}{Jiafeng Guo}, \bibinfo{person}{Yanyan Lan}, \bibinfo{person}{Jun Xu}, \bibinfo{person}{Chengxiang Zhai}, {and} \bibinfo{person}{Xueqi Cheng}.} \bibinfo{year}{2018}\natexlab{}.
\newblock \showarticletitle{Modeling Diverse Relevance Patterns in Ad-hoc Retrieval}. In \bibinfo{booktitle}{\emph{SIGIR}}. \bibinfo{pages}{375--384}.
\newblock


\bibitem[Fan et~al\mbox{.}(2022)]%
        {fan2022pre}
\bibfield{author}{\bibinfo{person}{Yixing Fan}, \bibinfo{person}{Xiaohui Xie}, \bibinfo{person}{Yinqiong Cai}, \bibinfo{person}{Jia Chen}, \bibinfo{person}{Xinyu Ma}, \bibinfo{person}{Xiangsheng Li}, \bibinfo{person}{Ruqing Zhang}, \bibinfo{person}{Jiafeng Guo}, {et~al\mbox{.}}} \bibinfo{year}{2022}\natexlab{}.
\newblock \showarticletitle{Pre-training Methods in Information Retrieval}.
\newblock \bibinfo{journal}{\emph{Foundations and Trends{\textregistered} in Information Retrieval}} \bibinfo{volume}{16}, \bibinfo{number}{3} (\bibinfo{year}{2022}), \bibinfo{pages}{178--317}.
\newblock


\bibitem[Fang et~al\mbox{.}(2023)]%
        {fang-etal-2023-modeling}
\bibfield{author}{\bibinfo{person}{Xuanjie Fang}, \bibinfo{person}{Sijie Cheng}, \bibinfo{person}{Yang Liu}, {and} \bibinfo{person}{Wei Wang}.} \bibinfo{year}{2023}\natexlab{}.
\newblock \showarticletitle{Modeling Adversarial Attack on Pre-trained Language Models as Sequential Decision Making}. In \bibinfo{booktitle}{\emph{Findings of the Association for Computational Linguistics}}. \bibinfo{pages}{7322--7336}.
\newblock


\bibitem[Gramamrly(2023)]%
        {grammarly}
\bibfield{author}{\bibinfo{person}{Gramamrly}.} \bibinfo{year}{2023}\natexlab{}.
\newblock \bibinfo{howpublished}{\url{https://app.grammarly.com/}}.
\newblock


\bibitem[Guo et~al\mbox{.}(2016)]%
        {guo2016deep}
\bibfield{author}{\bibinfo{person}{Jiafeng Guo}, \bibinfo{person}{Yixing Fan}, \bibinfo{person}{Qingyao Ai}, {and} \bibinfo{person}{W~Bruce Croft}.} \bibinfo{year}{2016}\natexlab{}.
\newblock \showarticletitle{A Deep Relevance Matching Model for Ad-hoc Retrieval}. In \bibinfo{booktitle}{\emph{CIKM}}. \bibinfo{pages}{55--64}.
\newblock


\bibitem[Gyongyi and Garcia-Molina(2005)]%
        {gyongyi2005web}
\bibfield{author}{\bibinfo{person}{Zoltan Gyongyi} {and} \bibinfo{person}{Hector Garcia-Molina}.} \bibinfo{year}{2005}\natexlab{}.
\newblock \showarticletitle{Web Spam Taxonomy}. In \bibinfo{booktitle}{\emph{AIRWeb}}.
\newblock


\bibitem[He et~al\mbox{.}(2020)]%
        {he2020deberta}
\bibfield{author}{\bibinfo{person}{Pengcheng He}, \bibinfo{person}{Xiaodong Liu}, \bibinfo{person}{Jianfeng Gao}, {and} \bibinfo{person}{Weizhu Chen}.} \bibinfo{year}{2020}\natexlab{}.
\newblock \showarticletitle{DeBERTa: Decoding-enhanced BERT with Disentangled Attention}. In \bibinfo{booktitle}{\emph{ICLR}}.
\newblock


\bibitem[He et~al\mbox{.}(2023)]%
        {he2023merging}
\bibfield{author}{\bibinfo{person}{Shwai He}, \bibinfo{person}{Run-Ze Fan}, \bibinfo{person}{Liang Ding}, \bibinfo{person}{Li Shen}, \bibinfo{person}{Tianyi Zhou}, {and} \bibinfo{person}{Dacheng Tao}.} \bibinfo{year}{2023}\natexlab{}.
\newblock \showarticletitle{Merging Experts into One: Improving Computational Efficiency of Mixture of Experts}. In \bibinfo{booktitle}{\emph{Proceedings of the 2023 Conference on Empirical Methods in Natural Language Processing}}. \bibinfo{pages}{14685--14691}.
\newblock


\bibitem[Jin et~al\mbox{.}(2023)]%
        {jin2023adversarial}
\bibfield{author}{\bibinfo{person}{Xin Jin}, \bibinfo{person}{Ruxin Wang}, \bibinfo{person}{Shin-Jye Lee}, \bibinfo{person}{Qian Jiang}, \bibinfo{person}{Shaowen Yao}, {and} \bibinfo{person}{Wei Zhou}.} \bibinfo{year}{2023}\natexlab{}.
\newblock \showarticletitle{Adversarial Attacks on Multi-focus Image Fusion Models}.
\newblock \bibinfo{journal}{\emph{Computers \& Security}}  \bibinfo{volume}{134} (\bibinfo{year}{2023}), \bibinfo{pages}{103455}.
\newblock


\bibitem[Kenton and Toutanova(2019)]%
        {devlin2018bert}
\bibfield{author}{\bibinfo{person}{Jacob Devlin Ming-Wei~Chang Kenton} {and} \bibinfo{person}{Lee~Kristina Toutanova}.} \bibinfo{year}{2019}\natexlab{}.
\newblock \showarticletitle{BERT: Pre-training of Deep Bidirectional Transformers for Language Understanding}. In \bibinfo{booktitle}{\emph{NAACL-HLT}}.
\newblock


\bibitem[Kocsis and Szepesv{\'a}ri(2006)]%
        {LeventeKocsis2006BanditBM}
\bibfield{author}{\bibinfo{person}{Levente Kocsis} {and} \bibinfo{person}{Csaba Szepesv{\'a}ri}.} \bibinfo{year}{2006}\natexlab{}.
\newblock \showarticletitle{Bandit Based Monte-Carlo Planning}. In \bibinfo{booktitle}{\emph{ECML}}.
\newblock


\bibitem[Kurland and Tennenholtz(2022)]%
        {kurland2022competitive}
\bibfield{author}{\bibinfo{person}{Oren Kurland} {and} \bibinfo{person}{Moshe Tennenholtz}.} \bibinfo{year}{2022}\natexlab{}.
\newblock \showarticletitle{Competitive Search}. In \bibinfo{booktitle}{\emph{SIGIR}}.
\newblock


\bibitem[Kwiatkowski et~al\mbox{.}(2019)]%
        {kwiatkowski2019natural}
\bibfield{author}{\bibinfo{person}{Tom Kwiatkowski}, \bibinfo{person}{Jennimaria Palomaki}, \bibinfo{person}{Olivia Redfield}, \bibinfo{person}{Michael Collins}, \bibinfo{person}{Ankur Parikh}, \bibinfo{person}{Chris Alberti}, \bibinfo{person}{Danielle Epstein}, \bibinfo{person}{Illia Polosukhin}, \bibinfo{person}{Jacob Devlin}, \bibinfo{person}{Kenton Lee}, {et~al\mbox{.}}} \bibinfo{year}{2019}\natexlab{}.
\newblock \showarticletitle{Natural questions: a benchmark for question answering research}.
\newblock \bibinfo{journal}{\emph{Transactions of the Association for Computational Linguistics}}  \bibinfo{volume}{7} (\bibinfo{year}{2019}), \bibinfo{pages}{453--466}.
\newblock


\bibitem[LeCun et~al\mbox{.}(2015)]%
        {lecun2015deep}
\bibfield{author}{\bibinfo{person}{Yann LeCun}, \bibinfo{person}{Yoshua Bengio}, {and} \bibinfo{person}{Geoffrey Hinton}.} \bibinfo{year}{2015}\natexlab{}.
\newblock \showarticletitle{Deep Learning}.
\newblock \bibinfo{journal}{\emph{Nature}} \bibinfo{volume}{521}, \bibinfo{number}{7553} (\bibinfo{year}{2015}), \bibinfo{pages}{436--444}.
\newblock


\bibitem[Lei et~al\mbox{.}(2022)]%
        {lei2022phrase}
\bibfield{author}{\bibinfo{person}{Yibin Lei}, \bibinfo{person}{Yu Cao}, \bibinfo{person}{Dianqi Li}, \bibinfo{person}{Tianyi Zhou}, \bibinfo{person}{Meng Fang}, {and} \bibinfo{person}{Mykola Pechenizkiy}.} \bibinfo{year}{2022}\natexlab{}.
\newblock \showarticletitle{Phrase-level Textual Adversarial Attack with Label Preservation}. In \bibinfo{booktitle}{\emph{Findings of the Association for Computational Linguistics: NAACL 2022}}. \bibinfo{pages}{1095--1112}.
\newblock


\bibitem[Lewis et~al\mbox{.}(2020)]%
        {lewis2020bart}
\bibfield{author}{\bibinfo{person}{Mike Lewis}, \bibinfo{person}{Yinhan Liu}, \bibinfo{person}{Naman Goyal}, \bibinfo{person}{Marjan Ghazvininejad}, \bibinfo{person}{Abdelrahman Mohamed}, \bibinfo{person}{Omer Levy}, \bibinfo{person}{Veselin Stoyanov}, {and} \bibinfo{person}{Luke Zettlemoyer}.} \bibinfo{year}{2020}\natexlab{}.
\newblock \showarticletitle{BART: Denoising Sequence-to-Sequence Pre-training for Natural Language Generation, Translation, and Comprehension}. In \bibinfo{booktitle}{\emph{ACL}}. \bibinfo{pages}{7871--7880}.
\newblock


\bibitem[Li(2014)]%
        {li2014learning}
\bibfield{author}{\bibinfo{person}{Hang Li}.} \bibinfo{year}{2014}\natexlab{}.
\newblock \showarticletitle{Learning to Rank for Information Retrieval and Natural Language Processing}.
\newblock \bibinfo{journal}{\emph{Synth. Lect. Hum. Lang. Technol.}} \bibinfo{volume}{7}, \bibinfo{number}{3} (\bibinfo{year}{2014}), \bibinfo{pages}{1--121}.
\newblock


\bibitem[Li et~al\mbox{.}(2019)]%
        {li2019textbugger}
\bibfield{author}{\bibinfo{person}{Jinfeng Li}, \bibinfo{person}{Shouling Ji}, \bibinfo{person}{Tianyu Du}, \bibinfo{person}{Bo Li}, {and} \bibinfo{person}{Ting Wang}.} \bibinfo{year}{2019}\natexlab{}.
\newblock \showarticletitle{TextBugger: Generating Adversarial Text Against Real-world Applications}. In \bibinfo{booktitle}{\emph{NDSS}}.
\newblock


\bibitem[Li et~al\mbox{.}(2023)]%
        {li2023generative}
\bibfield{author}{\bibinfo{person}{Junlong Li}, \bibinfo{person}{Shichao Sun}, \bibinfo{person}{Weizhe Yuan}, \bibinfo{person}{Run-Ze Fan}, \bibinfo{person}{Hai Zhao}, {and} \bibinfo{person}{Pengfei Liu}.} \bibinfo{year}{2023}\natexlab{}.
\newblock \showarticletitle{Generative Judge for Evaluating Alignment}.
\newblock \bibinfo{journal}{\emph{arXiv preprint arXiv:2310.05470}} (\bibinfo{year}{2023}).
\newblock


\bibitem[Liang et~al\mbox{.}(2022)]%
        {Liangke_Survey}
\bibfield{author}{\bibinfo{person}{Ke Liang}, \bibinfo{person}{Lingyuan Meng}, \bibinfo{person}{Meng Liu}, \bibinfo{person}{Yue Liu}, \bibinfo{person}{Wenxuan Tu}, \bibinfo{person}{Siwei Wang}, \bibinfo{person}{Sihang Zhou}, \bibinfo{person}{Xinwang Liu}, {and} \bibinfo{person}{Fuchun Sun}.} \bibinfo{year}{2022}\natexlab{}.
\newblock \showarticletitle{Reasoning over different types of knowledge graphs: Static, temporal and multi-modal}.
\newblock \bibinfo{journal}{\emph{arXiv preprint arXiv:2212.05767}} (\bibinfo{year}{2022}).
\newblock


\bibitem[Liang et~al\mbox{.}(2023)]%
        {liang2023message}
\bibfield{author}{\bibinfo{person}{Ke Liang}, \bibinfo{person}{Lingyuan Meng}, \bibinfo{person}{Sihang Zhou}, \bibinfo{person}{Siwei Wang}, \bibinfo{person}{Wenxuan Tu}, \bibinfo{person}{Yue Liu}, \bibinfo{person}{Meng Liu}, {and} \bibinfo{person}{Xinwang Liu}.} \bibinfo{year}{2023}\natexlab{}.
\newblock \showarticletitle{Message intercommunication for inductive relation reasoning}.
\newblock \bibinfo{journal}{\emph{arXiv preprint arXiv:2305.14074}} (\bibinfo{year}{2023}).
\newblock


\bibitem[Liu et~al\mbox{.}(2022)]%
        {liu2022order}
\bibfield{author}{\bibinfo{person}{Jiawei Liu}, \bibinfo{person}{Yangyang Kang}, \bibinfo{person}{Di Tang}, \bibinfo{person}{Kaisong Song}, \bibinfo{person}{Changlong Sun}, \bibinfo{person}{Xiaofeng Wang}, \bibinfo{person}{Wei Lu}, {and} \bibinfo{person}{Xiaozhong Liu}.} \bibinfo{year}{2022}\natexlab{}.
\newblock \showarticletitle{Order-Disorder: Imitation Adversarial Attacks for Black-box Neural Ranking Models}. In \bibinfo{booktitle}{\emph{CCS}}. \bibinfo{pages}{2025--2039}.
\newblock


\bibitem[Liu(2009)]%
        {TieYanLiu2009LearningTR}
\bibfield{author}{\bibinfo{person}{Tie-Yan Liu}.} \bibinfo{year}{2009}\natexlab{}.
\newblock \showarticletitle{Learning to Rank for Information Retrieval}.
\newblock \bibinfo{journal}{\emph{Foundations and Trends in Information Retrieval}} \bibinfo{volume}{3}, \bibinfo{number}{3} (\bibinfo{year}{2009}), \bibinfo{pages}{225--331}.
\newblock


\bibitem[Liu et~al\mbox{.}(2023a)]%
        {liu2023robustness}
\bibfield{author}{\bibinfo{person}{Yu-An Liu}, \bibinfo{person}{Ruqing Zhang}, \bibinfo{person}{Jiafeng Guo}, \bibinfo{person}{Wei Chen}, {and} \bibinfo{person}{Xueqi Cheng}.} \bibinfo{year}{2023}\natexlab{a}.
\newblock \showarticletitle{On the Robustness of Generative Retrieval Models: An Out-of-Distribution Perspective}. In \bibinfo{booktitle}{\emph{Gen-IR@SIGIR}}.
\newblock


\bibitem[Liu et~al\mbox{.}(2023b)]%
        {liu2023black}
\bibfield{author}{\bibinfo{person}{Yu-An Liu}, \bibinfo{person}{Ruqing Zhang}, \bibinfo{person}{Jiafeng Guo}, \bibinfo{person}{Maarten de Rijke}, \bibinfo{person}{Wei Chen}, \bibinfo{person}{Yixing Fan}, {and} \bibinfo{person}{Xueqi Cheng}.} \bibinfo{year}{2023}\natexlab{b}.
\newblock \showarticletitle{Black-Box Adversarial Attacks against Dense Retrieval Models: A Multi-View Contrastive Learning Method}. In \bibinfo{booktitle}{\emph{CIKM}}. \bibinfo{pages}{1647–1656}.
\newblock


\bibitem[Liu et~al\mbox{.}(2023c)]%
        {liu2023topic}
\bibfield{author}{\bibinfo{person}{Yu-An Liu}, \bibinfo{person}{Ruqing Zhang}, \bibinfo{person}{Jiafeng Guo}, \bibinfo{person}{Maarten de Rijke}, \bibinfo{person}{Wei Chen}, \bibinfo{person}{Yixing Fan}, {and} \bibinfo{person}{Xueqi Cheng}.} \bibinfo{year}{2023}\natexlab{c}.
\newblock \showarticletitle{Topic-Oriented Adversarial Attacks against Black-Box Neural Ranking Models}. In \bibinfo{booktitle}{\emph{SIGIR}}. \bibinfo{pages}{1700–1709}.
\newblock


\bibitem[Liu et~al\mbox{.}(2024)]%
        {liu2024perturbation}
\bibfield{author}{\bibinfo{person}{Yu-An Liu}, \bibinfo{person}{Ruqing Zhang}, \bibinfo{person}{Mingkun Zhang}, \bibinfo{person}{Wei Chen}, \bibinfo{person}{Maarten de Rijke}, \bibinfo{person}{Jiafeng Guo}, {and} \bibinfo{person}{Xueqi Cheng}.} \bibinfo{year}{2024}\natexlab{}.
\newblock \showarticletitle{Perturbation-Invariant Adversarial Training for Neural Ranking Models: Improving the Effectiveness-Robustness Trade-Off}. In \bibinfo{booktitle}{\emph{Proceedings of the AAAI Conference on Artificial Intelligence}}, Vol.~\bibinfo{volume}{38}.
\newblock


\bibitem[Ma et~al\mbox{.}(2021)]%
        {ma2021prop}
\bibfield{author}{\bibinfo{person}{Xinyu Ma}, \bibinfo{person}{Jiafeng Guo}, \bibinfo{person}{Ruqing Zhang}, \bibinfo{person}{Yixing Fan}, \bibinfo{person}{Xiang Ji}, {and} \bibinfo{person}{Xueqi Cheng}.} \bibinfo{year}{2021}\natexlab{}.
\newblock \showarticletitle{Prop: Pre-training with Representative Words Prediction for Ad-hoc Retrieval}. In \bibinfo{booktitle}{\emph{WSDM}}. \bibinfo{pages}{283--291}.
\newblock


\bibitem[Mazyavkina et~al\mbox{.}(2021)]%
        {mazyavkina2021reinforcement}
\bibfield{author}{\bibinfo{person}{Nina Mazyavkina}, \bibinfo{person}{Sergey Sviridov}, \bibinfo{person}{Sergei Ivanov}, {and} \bibinfo{person}{Evgeny Burnaev}.} \bibinfo{year}{2021}\natexlab{}.
\newblock \showarticletitle{Reinforcement learning for combinatorial optimization: A survey}.
\newblock \bibinfo{journal}{\emph{Computers \& Operations Research}}  \bibinfo{volume}{134} (\bibinfo{year}{2021}), \bibinfo{pages}{105400}.
\newblock


\bibitem[Mitra et~al\mbox{.}(2017)]%
        {mitra2017learning}
\bibfield{author}{\bibinfo{person}{Bhaskar Mitra}, \bibinfo{person}{Fernando Diaz}, {and} \bibinfo{person}{Nick Craswell}.} \bibinfo{year}{2017}\natexlab{}.
\newblock \showarticletitle{Learning to Match Using Local and Distributed Representations of Text for Web Search}. In \bibinfo{booktitle}{\emph{WWW}}.
\newblock


\bibitem[Nguyen et~al\mbox{.}(2017)]%
        {nguyen2017collective}
\bibfield{author}{\bibinfo{person}{Duc~Thien Nguyen}, \bibinfo{person}{Akshat Kumar}, {and} \bibinfo{person}{Hoong~Chuin Lau}.} \bibinfo{year}{2017}\natexlab{}.
\newblock \showarticletitle{Collective multiagent sequential decision making under uncertainty}. In \bibinfo{booktitle}{\emph{Proceedings of the AAAI Conference on Artificial Intelligence}}, Vol.~\bibinfo{volume}{31}.
\newblock


\bibitem[Nguyen et~al\mbox{.}(2016)]%
        {nguyen2016ms}
\bibfield{author}{\bibinfo{person}{Tri Nguyen}, \bibinfo{person}{Mir Rosenberg}, \bibinfo{person}{Xia Song}, \bibinfo{person}{Jianfeng Gao}, \bibinfo{person}{Saurabh Tiwary}, \bibinfo{person}{Rangan Majumder}, {and} \bibinfo{person}{Li Deng}.} \bibinfo{year}{2016}\natexlab{}.
\newblock \showarticletitle{MS MARCO: A Human Generated Machine Reading Comprehension Dataset}. In \bibinfo{booktitle}{\emph{CoCo@NIPS}}.
\newblock


\bibitem[Ni et~al\mbox{.}(2024)]%
        {ni2024llms}
\bibfield{author}{\bibinfo{person}{Shiyu Ni}, \bibinfo{person}{Keping Bi}, \bibinfo{person}{Jiafeng Guo}, {and} \bibinfo{person}{Xueqi Cheng}.} \bibinfo{year}{2024}\natexlab{}.
\newblock \showarticletitle{When Do LLMs Need Retrieval Augmentation? Mitigating LLMs' Overconfidence Helps Retrieval Augmentation}.
\newblock \bibinfo{journal}{\emph{arXiv preprint arXiv:2402.11457}} (\bibinfo{year}{2024}).
\newblock


\bibitem[Nogueira and Cho(2019)]%
        {RodrigoNogueira2019PassageRW}
\bibfield{author}{\bibinfo{person}{Rodrigo Nogueira} {and} \bibinfo{person}{Kyunghyun Cho}.} \bibinfo{year}{2019}\natexlab{}.
\newblock \showarticletitle{Passage Re-ranking with BERT}.
\newblock \bibinfo{journal}{\emph{arXiv preprint arXiv:1901.04085}} (\bibinfo{year}{2019}).
\newblock


\bibitem[Onal et~al\mbox{.}(2018)]%
        {KezbanDilekOnal2018NeuralIR}
\bibfield{author}{\bibinfo{person}{Kezban~Dilek Onal}, \bibinfo{person}{Ye Zhang}, \bibinfo{person}{Ismail~Sengor Altingovde}, \bibinfo{person}{Md.~Mustafizur Rahman}, \bibinfo{person}{Pinar Karagoz}, \bibinfo{person}{Alexander Braylan}, \bibinfo{person}{Brandon Dang}, \bibinfo{person}{Heng-Lu Chang}, \bibinfo{person}{Henna Kim}, \bibinfo{person}{Quinten McNamara}, \bibinfo{person}{Aaron Angert}, \bibinfo{person}{Edward Banner}, \bibinfo{person}{Vivek Khetan}, \bibinfo{person}{Tyler McDonnell}, \bibinfo{person}{An~Thanh Nguyen}, \bibinfo{person}{Dan Xu}, \bibinfo{person}{Byron~C. Wallace}, \bibinfo{person}{Maarten de Rijke}, {and} \bibinfo{person}{Matthew Lease}.} \bibinfo{year}{2018}\natexlab{}.
\newblock \showarticletitle{Neural Information Retrieval: At the End of the Early Years}.
\newblock \bibinfo{journal}{\emph{Information Retrieval}} \bibinfo{volume}{21}, \bibinfo{number}{2–3} (\bibinfo{year}{2018}), \bibinfo{pages}{111--182}.
\newblock


\bibitem[OpenAI(2022)]%
        {chatgpt}
\bibfield{author}{\bibinfo{person}{OpenAI}.} \bibinfo{year}{2022}\natexlab{}.
\newblock \bibinfo{title}{Introducing ChatGPT}.
\newblock \bibinfo{howpublished}{\url{https://openai.com/blog/chatgpt}}.
\newblock


\bibitem[Ponte and Croft(1998)]%
        {JayPonte1998ALM}
\bibfield{author}{\bibinfo{person}{Jay Ponte} {and} \bibinfo{person}{W.~Bruce Croft}.} \bibinfo{year}{1998}\natexlab{}.
\newblock \showarticletitle{A Language Modeling Approach to Information Retrieval}.
\newblock \bibinfo{journal}{\emph{SIGIR}}.
\newblock


\bibitem[Pruthi et~al\mbox{.}(2019)]%
        {pruthi2019combating}
\bibfield{author}{\bibinfo{person}{Danish Pruthi}, \bibinfo{person}{Bhuwan Dhingra}, {and} \bibinfo{person}{Zachary~C Lipton}.} \bibinfo{year}{2019}\natexlab{}.
\newblock \showarticletitle{Combating Adversarial Misspellings with Robust Word Recognition}. In \bibinfo{booktitle}{\emph{ACL}}. \bibinfo{pages}{5582--5591}.
\newblock


\bibitem[Radford et~al\mbox{.}(2019)]%
        {radford2019language}
\bibfield{author}{\bibinfo{person}{Alec Radford}, \bibinfo{person}{Jeffrey Wu}, \bibinfo{person}{Rewon Child}, \bibinfo{person}{David Luan}, \bibinfo{person}{Dario Amodei}, {and} \bibinfo{person}{Ilya Sutskever}.} \bibinfo{year}{2019}\natexlab{}.
\newblock \showarticletitle{Language Models Are Unsupervised Multitask Learners}.
\newblock \bibinfo{journal}{\emph{OpenAI blog}} \bibinfo{volume}{1}, \bibinfo{number}{8} (\bibinfo{year}{2019}), \bibinfo{pages}{9}.
\newblock


\bibitem[Raval and Verma(2020)]%
        {raval2020one}
\bibfield{author}{\bibinfo{person}{Nisarg Raval} {and} \bibinfo{person}{Manisha Verma}.} \bibinfo{year}{2020}\natexlab{}.
\newblock \showarticletitle{One Word at a Time: Adversarial Attacks on Retrieval Models}.
\newblock \bibinfo{journal}{\emph{arXiv preprint arXiv:2008.02197}} (\bibinfo{year}{2020}).
\newblock


\bibitem[Robertson and Walker(1994)]%
        {StephenRobertson1994SomeSE}
\bibfield{author}{\bibinfo{person}{Stephen~E Robertson} {and} \bibinfo{person}{Steve Walker}.} \bibinfo{year}{1994}\natexlab{}.
\newblock \showarticletitle{Some Simple Effective Approximations to the 2-poisson Model for Probabilistic Weighted Retrieval}. In \bibinfo{booktitle}{\emph{SIGIR’94}}. Springer, \bibinfo{pages}{232--241}.
\newblock


\bibitem[Rumelhart et~al\mbox{.}(1985)]%
        {rumelhart1985learning}
\bibfield{author}{\bibinfo{person}{David~E Rumelhart}, \bibinfo{person}{Geoffrey~E Hinton}, {and} \bibinfo{person}{Ronald~J Williams}.} \bibinfo{year}{1985}\natexlab{}.
\newblock \bibinfo{booktitle}{\emph{Learning Internal Representations by Error Propagation}}.
\newblock \bibinfo{type}{{T}echnical {R}eport}. \bibinfo{institution}{California Univ San Diego La Jolla Inst for Cognitive Science}.
\newblock


\bibitem[Salton et~al\mbox{.}(1975)]%
        {GerardSalton1975AVS}
\bibfield{author}{\bibinfo{person}{Gerard Salton}, \bibinfo{person}{A. Wong}, {and} \bibinfo{person}{C.~S. Yang}.} \bibinfo{year}{1975}\natexlab{}.
\newblock \showarticletitle{A Vector Space Model for Automatic Indexing}.
\newblock \bibinfo{journal}{\emph{Commun. ACM}} \bibinfo{volume}{18}, \bibinfo{number}{11} (\bibinfo{year}{1975}), \bibinfo{pages}{613–620}.
\newblock


\bibitem[Samanta and Mehta(2017)]%
        {samanta2017towards}
\bibfield{author}{\bibinfo{person}{Suranjana Samanta} {and} \bibinfo{person}{Sameep Mehta}.} \bibinfo{year}{2017}\natexlab{}.
\newblock \showarticletitle{Towards Crafting Text Adversarial Samples}.
\newblock \bibinfo{journal}{\emph{arXiv preprint arXiv:1707.02812}} (\bibinfo{year}{2017}).
\newblock


\bibitem[Song et~al\mbox{.}(2020)]%
        {CongzhengSong2020AdversarialSC}
\bibfield{author}{\bibinfo{person}{Congzheng Song}, \bibinfo{person}{Alexander~M. Rush}, {and} \bibinfo{person}{Vitaly Shmatikov}.} \bibinfo{year}{2020}\natexlab{}.
\newblock \showarticletitle{Adversarial Semantic Collisions}. In \bibinfo{booktitle}{\emph{EMNLP}}.
\newblock


\bibitem[Song et~al\mbox{.}(2022)]%
        {song2022trattack}
\bibfield{author}{\bibinfo{person}{Junshuai Song}, \bibinfo{person}{Jiangshan Zhang}, \bibinfo{person}{Jifeng Zhu}, \bibinfo{person}{Mengyun Tang}, {and} \bibinfo{person}{Yong Yang}.} \bibinfo{year}{2022}\natexlab{}.
\newblock \showarticletitle{TRAttack: Text Rewriting Attack against Text Retrieval}. In \bibinfo{booktitle}{\emph{Proceedings of the 7th Workshop on Representation Learning for NLP}}. \bibinfo{pages}{191--203}.
\newblock


\bibitem[Sun et~al\mbox{.}(2023)]%
        {sun-etal-2023-chatgpt}
\bibfield{author}{\bibinfo{person}{Weiwei Sun}, \bibinfo{person}{Lingyong Yan}, \bibinfo{person}{Xinyu Ma}, \bibinfo{person}{Shuaiqiang Wang}, \bibinfo{person}{Pengjie Ren}, \bibinfo{person}{Zhumin Chen}, \bibinfo{person}{Dawei Yin}, {and} \bibinfo{person}{Zhaochun Ren}.} \bibinfo{year}{2023}\natexlab{}.
\newblock \showarticletitle{Is {C}hat{GPT} Good at Search? Investigating Large Language Models as Re-Ranking Agents}. In \bibinfo{booktitle}{\emph{EMNLP}}. \bibinfo{pages}{14918--14937}.
\newblock


\bibitem[Sutton and Barto(2018)]%
        {sutton2018reinforcement}
\bibfield{author}{\bibinfo{person}{Richard~S. Sutton} {and} \bibinfo{person}{Andrew~G. Barto}.} \bibinfo{year}{2018}\natexlab{}.
\newblock \bibinfo{booktitle}{\emph{Reinforcement Learning: An Introduction}}.
\newblock \bibinfo{publisher}{MIT Press}.
\newblock


\bibitem[Tsai et~al\mbox{.}(2019)]%
        {tsai2019small}
\bibfield{author}{\bibinfo{person}{Henry Tsai}, \bibinfo{person}{Jason Riesa}, \bibinfo{person}{Melvin Johnson}, \bibinfo{person}{Naveen Arivazhagan}, \bibinfo{person}{Xin Li}, {and} \bibinfo{person}{Amelia Archer}.} \bibinfo{year}{2019}\natexlab{}.
\newblock \showarticletitle{Small and Practical BERT Models for Sequence Labeling}. In \bibinfo{booktitle}{\emph{EMNLP}}. \bibinfo{pages}{3632--3636}.
\newblock


\bibitem[Wang et~al\mbox{.}(2022a)]%
        {wang2022micro}
\bibfield{author}{\bibinfo{person}{Shaokun Wang}, \bibinfo{person}{Tian Gan}, \bibinfo{person}{Yuan Liu}, \bibinfo{person}{Jianlong Wu}, \bibinfo{person}{Yuan Cheng}, {and} \bibinfo{person}{Liqiang Nie}.} \bibinfo{year}{2022}\natexlab{a}.
\newblock \showarticletitle{Micro-influencer recommendation by multi-perspective account representation learning}.
\newblock \bibinfo{journal}{\emph{IEEE Transactions on Multimedia}} (\bibinfo{year}{2022}).
\newblock


\bibitem[Wang et~al\mbox{.}(2021)]%
        {wang2021discover}
\bibfield{author}{\bibinfo{person}{Shaokun Wang}, \bibinfo{person}{Tian Gan}, \bibinfo{person}{Yuan Liu}, \bibinfo{person}{Li Zhang}, \bibinfo{person}{JianLong Wu}, {and} \bibinfo{person}{Liqiang Nie}.} \bibinfo{year}{2021}\natexlab{}.
\newblock \showarticletitle{Discover micro-influencers for brands via better understanding}.
\newblock \bibinfo{journal}{\emph{IEEE Transactions on Multimedia}}  \bibinfo{volume}{24} (\bibinfo{year}{2021}), \bibinfo{pages}{2595--2605}.
\newblock


\bibitem[Wang et~al\mbox{.}(2022b)]%
        {wang2022bert}
\bibfield{author}{\bibinfo{person}{Yumeng Wang}, \bibinfo{person}{Lijun Lyu}, {and} \bibinfo{person}{Avishek Anand}.} \bibinfo{year}{2022}\natexlab{b}.
\newblock \showarticletitle{BERT Rankers are Brittle: A Study using Adversarial Document Perturbations}. In \bibinfo{booktitle}{\emph{ICTIR}}.
\newblock


\bibitem[Wu et~al\mbox{.}(2023b)]%
        {wu2022prada}
\bibfield{author}{\bibinfo{person}{Chen Wu}, \bibinfo{person}{Ruqing Zhang}, \bibinfo{person}{Jiafeng Guo}, \bibinfo{person}{Maarten de Rijke}, \bibinfo{person}{Yixing Fan}, {and} \bibinfo{person}{Xueqi Cheng}.} \bibinfo{year}{2023}\natexlab{b}.
\newblock \showarticletitle{PRADA: Practical Black-Box Adversarial Attacks against Neural Ranking Models}.
\newblock \bibinfo{journal}{\emph{TOIS}} \bibinfo{volume}{41}, \bibinfo{number}{4} (\bibinfo{year}{2023}), \bibinfo{pages}{Article 89}.
\newblock


\bibitem[Wu et~al\mbox{.}(2023a)]%
        {wu2023imperceptible}
\bibfield{author}{\bibinfo{person}{Guoming Wu}, \bibinfo{person}{Yangfan Xu}, \bibinfo{person}{Jun Li}, \bibinfo{person}{Zhiping Shi}, {and} \bibinfo{person}{Xianglong Liu}.} \bibinfo{year}{2023}\natexlab{a}.
\newblock \showarticletitle{Imperceptible Adversarial Attack with Multi-granular Spatio-temporal Attention for Video Action Recognition}.
\newblock \bibinfo{journal}{\emph{IEEE Internet of Things Journal}} (\bibinfo{year}{2023}).
\newblock


\bibitem[Yang et~al\mbox{.}(2018)]%
        {PeilinYang2018AnseriniRR}
\bibfield{author}{\bibinfo{person}{Peilin Yang}, \bibinfo{person}{Hui Fang}, {and} \bibinfo{person}{Jimmy Lin}.} \bibinfo{year}{2018}\natexlab{}.
\newblock \showarticletitle{Anserini: Reproducible Ranking Baselines Using Lucene}.
\newblock \bibinfo{journal}{\emph{Journal of Data and Information Quality}} \bibinfo{volume}{10}, \bibinfo{number}{4} (\bibinfo{year}{2018}), \bibinfo{pages}{Article 16}.
\newblock


\bibitem[Zhang et~al\mbox{.}(2023)]%
        {zhang2023relevance}
\bibfield{author}{\bibinfo{person}{Hengran Zhang}, \bibinfo{person}{Ruqing Zhang}, \bibinfo{person}{Jiafeng Guo}, \bibinfo{person}{Maarten de Rijke}, \bibinfo{person}{Yixing Fan}, {and} \bibinfo{person}{Xueqi Cheng}.} \bibinfo{year}{2023}\natexlab{}.
\newblock \showarticletitle{From Relevance to Utility: Evidence Retrieval with Feedback for Fact Verification}. In \bibinfo{booktitle}{\emph{Findings of the Association for Computational Linguistics: EMNLP 2023}}. \bibinfo{pages}{6373--6384}.
\newblock


\bibitem[Zhang et~al\mbox{.}(2020)]%
        {zhang2020adversarial}
\bibfield{author}{\bibinfo{person}{Wei~Emma Zhang}, \bibinfo{person}{Quan~Z Sheng}, \bibinfo{person}{Ahoud Alhazmi}, {and} \bibinfo{person}{Chenliang Li}.} \bibinfo{year}{2020}\natexlab{}.
\newblock \showarticletitle{Adversarial Attacks on Deep Learning Models in Natural Language Processing: A Survey}.
\newblock \bibinfo{journal}{\emph{ACM TIST}} \bibinfo{volume}{11}, \bibinfo{number}{3} (\bibinfo{year}{2020}), \bibinfo{pages}{1--41}.
\newblock


\bibitem[Zheng et~al\mbox{.}(2020)]%
        {zheng2020evaluating}
\bibfield{author}{\bibinfo{person}{Xiaoqing Zheng}, \bibinfo{person}{Jiehang Zeng}, \bibinfo{person}{Yi Zhou}, \bibinfo{person}{Cho-Jui Hsieh}, \bibinfo{person}{Minhao Cheng}, {and} \bibinfo{person}{Xuan-Jing Huang}.} \bibinfo{year}{2020}\natexlab{}.
\newblock \showarticletitle{Evaluating and Enhancing the Robustness of Neural Network-based Dependency Parsing Models with Adversarial Examples}. In \bibinfo{booktitle}{\emph{ACL}}. \bibinfo{pages}{6600--6610}.
\newblock


\bibitem[Zhou and Pei(2009)]%
        {zhou2009osd}
\bibfield{author}{\bibinfo{person}{Bin Zhou} {and} \bibinfo{person}{Jian Pei}.} \bibinfo{year}{2009}\natexlab{}.
\newblock \showarticletitle{OSD: An Online Web Spam Detection System}. In \bibinfo{booktitle}{\emph{In Proceedings of the 15th ACM SIGKDD International Conference on Knowledge Discovery and Data Mining, KDD}}, Vol.~\bibinfo{volume}{9}.
\newblock


\bibitem[Zou et~al\mbox{.}(2022)]%
        {zou2022automatic}
\bibfield{author}{\bibinfo{person}{Xinyu Zou}, \bibinfo{person}{Zhi Hu}, \bibinfo{person}{Yiming Zhao}, \bibinfo{person}{Xuchu Ding}, \bibinfo{person}{Zhongyi Liu}, \bibinfo{person}{Chenliang Li}, {and} \bibinfo{person}{Aixin Sun}.} \bibinfo{year}{2022}\natexlab{}.
\newblock \showarticletitle{Automatic Expert Selection for Multi-Scenario and Multi-Task Search}.
\newblock \bibinfo{journal}{\emph{arXiv preprint arXiv:2205.14321}} (\bibinfo{year}{2022}).
\newblock


\end{thebibliography}

\end{document}